\documentclass[11pt]{article}
\usepackage{array}
\usepackage{longtable}
\usepackage{graphicx}
\usepackage{listings}

\usepackage{fancyhdr}
\usepackage{graphicx}
\usepackage{amssymb}
\usepackage{epstopdf}
\usepackage{amsmath} 	
\usepackage{amssymb}
\usepackage{cite}
\usepackage{multirow}
\usepackage[table]{xcolor}
\usepackage{wrapfig}
\usepackage{subfig}
\usepackage{todonotes}
\usepackage{listings}
\usepackage{wasysym}
\usepackage{float} % For Figure placement
\usepackage[colorlinks=true, linkcolor=black,citecolor=black,urlcolor=blue]{hyperref}
\usepackage{soul}
\setul{0.5ex}{0.3ex}
\definecolor{Blue}{rgb}{0,0.0,1}
\setulcolor{Blue} 

\usepackage[utf8]{inputenc}
\usepackage[T1]{fontenc}
\usepackage{lmodern}

\usepackage{graphicx}

\usepackage{array}
\usepackage{longtable}

\usepackage{amsmath, amssymb}

\usepackage{hyperref} % For hyperlinks
\usepackage{xcolor} % For color

\usepackage[a4paper, margin=1in]{geometry}

\usepackage{setspace}
\usepackage{times}

\begin{document}
%\title{{\color{red} Project guidelines and suggested template}}
\thispagestyle{empty}

\title{MLOPS in a multicloud environment: Typical Network Topology}
\author{Boyang Yan\thanks{byan4@ncsu.edu NC State University, Computer Science, USA}}

%%%\subtitle{Untertitel / Subtitle} % if needed
%\author[1]{Boyang Yan}{byan4@ncsu.edu}{0000-0002-8546-5004}
%\author[2]{Firstname2 Lastname2}{vorname.name@affiliation2.de}{0000-0000-0000-0000}
%\author[1]{Firstname3 Lastname 3}{vorname.name@affiliation1.de}{0000-0000-0000-0000}
%\author[1]{Firstname4 Lastname 4}{vorname.name@affiliation1.de}{0000-0000-0000-0000}
%\affil[1]{NC State University\\Computer Science\\USA}
%\affil[2]{University\\Department\\Address\\Country}

\maketitle

\begin{abstract}
As artificial intelligence, machine learning, and data science continue to drive the data-centric economy, the challenges of implementing machine learning on a single machine due to extensive data and computational needs have led to the adoption of cloud computing solutions. This research paper explores the design and implementation of a secure, cloud-native machine learning operations (MLOPS) pipeline that supports multi-cloud environments. The primary objective is to create a robust infrastructure that facilitates secure data collection, real-time model inference, and efficient management of the machine learning lifecycle. By leveraging cloud providers' capabilities, the solution aims to streamline the deployment and maintenance of machine learning models, ensuring high availability, scalability, and security. This paper details the network topology, problem description, business and technical requirements, trade-offs, and the provider selection process for achieving an optimal MLOPS environment.
\end{abstract}

%%%%%%%%%%%%%%%%%%%%%%%%%%%%%%%%%%%%%%%%%%%%%%%%%%%%%%%%%%%%%%%%
%%%%%%%%%%%%%%%%%%%%%%%%%%%%%%%%%%%%%%%%%%%%%%%%%%%%%%%%%%%%%%%%

% \input{0guidelines}

\section{Introduction}
\label{sec:intro}

\subsection{Motivation}

Artificial intelligence, machine learning, and data science are becoming more and more important as the world moves toward a more data-driven economy.
Machine learning is challenging to implement on a single computer, because to the vast amount of data and the requirement for computational power. As a result, cloud computing has emerged as the most effective tool for resolving these two problems.

Our project aims to make a secure network and privileged, isolated environment, multi-team support, and lightweight machine learning life cycle solution to handle any generic machine learning task and speed up the delivery of actionable results that enable better performance.

% {\color{red} Describe your motivation in doing this project\footnote{Here is an example of a valid motivation: I had no choice, the professors made me do it.}.}

\subsection{Executive summary}
% {\color{red} Summarize, in one paragraph or two, the essence of this project.} (Who are you writing this summary for?)

The project's primary goal is to design a cloud-native machine learning operations (MLOPS) pipeline. Cloud-based machine learning operation is a set of practices that manage and maintain the machine learning life cycle via pipeline in production reliably and efficiently hosted on the cloud provider.

Our target readers of the project summary are university research groups and individuals who want to use cloud provider resources to discover the possibility of automating the machine learning pipeline on the cloud and applying machine learning and deep learning methods within research with cloud-based machine learning architecture solutions.

We aim to provide a cloud-hosted solution that handles data collection, maintenance of training clusters, setup of the evaluation environment for machine learning, real-time and nearly real-time interference of models, and model monitoring for re-training. Our solution's standout feature allows for different individuals or organizations to take part in a single research endeavor while offering a variety of security measures. The completed trained model can be immediately distributed to edge devices by web searches, and the research data will not be exposed.

Due to the large number of people and organizations involved, the machine learning algorithm's quick iteration cycle, and model deployment, managing the entire machine learning life cycle becomes difficult.

\newpage
\section{Problem Description}
\label{sec:Problem Description}
  
\subsection{The problem}

For the cloud computing application, the specific scenario would be to automate the functioning of a machine learning system to detect objects and classify them into a few categories. 

Understanding two crucial principles for the machine learning life cycle across a cloud-based machine learning architecture pipeline is vitally necessary for understanding cloud architecture.

% {\color{red} What is the high-level problem you are designing for?} Must be a realistic one, in order to derive good Business Requirements (BRs).

% Potential suggestions for topics, taken from \\ \url{https://pg-p.ctme.caltech.edu/cloud-computing-certification-course#form-id}.\\
% The first four come from a list of projects students in Caltech are working on as part of the online MS degree. We added a brief description of the sixth topic in order to give you an idea. 

%% YP: We can add a "Online Bookstore".
%% YV: added
%%YV: We will not provide code for these projects. The idea here is to help the students come up with a realistic problem, from which they will derive good requirements (along the lines of the Australia example.

\begin{enumerate}
    \item \textbf{Data flow} The project will require an on-premises network to cloud network topology because dataflow concerning how to manage and store the data and how to move the data in the machine learning pipeline while moving data is highly related Networking.
    
    \item \textbf{Compute Resources/Environments} about how to compute the data on each step; different steps to run in different environments.
\end{enumerate}

Based on the above two principles for the machine learning life cycle, the whole system is classified two parts for sourcing the data, \textbf{On-premise Design}, and \textbf{Cloud Design}.

\begin{figure}[htp]
    \centering
    \includegraphics[width=15cm]{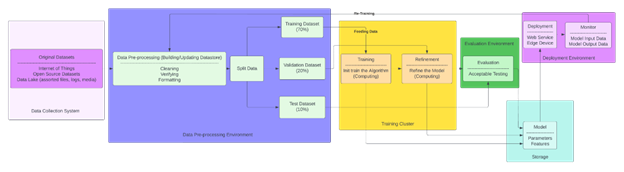}
    % \caption{Cloud architecture diagram hosting the MLOPS application}
    \label{fig:galaxy}
\end{figure}

The On-premise/Edge design has two sub-system.

\begin{enumerate}
    \item \textbf{Data Collection System} 
    
    Gathering data is one of the most important stages of machine learning workflows. Most of the original datasets need to be collected from the on-premise network. Aggregating multi-sensor data will be faced multiple problems, such as time synchronization of multi-sensor with cloud data centers. In this sub-system, we focus on solving time synchronization issues and collecting data from multiple sensors into the data lake of the cloud.
    
    \item \textbf{Machine Learning Model Real-Time Interference}

    When a new version of the trained ML model is done in the cloud, in order to on-premise/edge users could have a real-time interference experience. We are designing a new model distribution system from the cloud to an on-premise gateway or edge devices.

\end{enumerate}

The Cloud design can be broadly classified into five main sub-systems. 

\begin{enumerate}
    \item \textbf{Data Preprocessing} 
    
Many widely used frameworks and libraries will be used for performing data preparation. In order to create a standard base Data Preprocessing environment, Ansible will manage this portion of the environment.
    
    \item \textbf{Training Cluster}

The most expensive component is the training cluster. This component of our system must be built as required and made available for use without delay after training.

    \item \textbf{Evaluation Environment} 
        
    \item \textbf{Cloud Model Near-Real time Inference Environment}

In the cloud data center, inference can also be done; managing quickly iterating models and model monitoring for re-training are two major issues.
    \item \textbf{Storage (Datastory and ML Model Storage)}

Data includes structured and unstructured types, and storage the diversity of data types is a significant difficulty.

\textbf{Note}: Datastory and Datasets are two different concepts. Datastory involves different datasets, such as Training Dataset, Validation Dataset, and Test Dataset etc.

\end{enumerate}

Although these subsystems are largely considered as standards, they should be adaptable to different machine-learning scenarios by changing their methodologies.

Inorder to restrict the scope of the project, we assume the input images are provided from an on-premise employed system (mostly a camera or sensor attached device that captures the image and transfers it into the pipeline). Once it reaches the pipeline, we broadly perform the following operations:

\begin{enumerate}
    \item \textbf{Data Preprocessing Service} - This service receives the input data from the data collection system. Once the image data is received, it needs to be cleaned (filtered for resolution and reorganizing the orientations), filtered, and formatted. These preprocessed images need to be stored for periodic retraining of the models. Then the data has to be split into training, validation, and test data.

    \item \textbf{Training Service} - The service trains the model using the preprocessed data. The algorithm would require tuning the hyper-parameters, which are done using the validation data set.

    \item \textbf{Evaluation Service} - This service uses the refined model and predicts the output of the test data set.
        
    \item \textbf{Storage Service} - This service stores the pre-processed images, pickles the models with weights and hyper-parameters, and stores the output results.

    \item \textbf{Deployment Service} - This service receives the output from the evaluation service and models it into appropriate output data. The output data is then sent to a web service that displays the results.

\end{enumerate}

\subsection{Business Requirements} 
The business requirements for the system handling our application are detailed below:
\begin{enumerate}
\item \textbf{BR1}: The system should be able to accept very large sizes of data from the on-premise data provider as input.
\item \textbf{BR2}: The system should be able to store very large volumes of data.
\item \textbf{BR3}: The system should be able to accept very high volumes of requests from the data provider.
\item \textbf{BR4}: The system should be scalable with new types of data input.
\item \textbf{BR5}: The system should allow consumers (subject to data residency restrictions) to store the data in a data center within their jurisdiction.
\item \textbf {BR6}: The cloud system should be more cost-efficient than hosting on-premise services.
\item \textbf {BR7}: The maintenance cost of the system should be affordable.
\item \textbf {BR8}: Human effort required to maintain the system should be minimal.
\item \textbf {BR9}: The system should have high availability.
\item \textbf {BR10}: The system should have a very high throughput.
\item \textbf {BR11}: The system should have no latency with respect to performance.
\item \textbf {BR12}: The system should have high resilience and a quick recovery time. 
\item \textbf {BR13}: The system should have a real-time backup for storing the data and the business application.
\item \textbf {BR14}: The system should be capable of rapidly deploying the requested services.
\item \textbf {BR15}: The system should be able to scale up and down based on the amount of load on the applications.
\item \textbf {BR24}: The system should isolate components such that different components cannot interfere with each other.
\item \textbf {BR16}: The system should be able to identify and distinguish the requests from each user (tenant) uniquely.
\item \textbf {BR17}: The system should ensure the confidentiality of information.
\item \textbf {BR18}: The system should ensure data security.
\item \textbf {BR19}: The system should provide a hack-free environment.
\item \textbf {BR20}: The system should prevent external services from accessing system-sensitive information.
\item \textbf {BR21}: The system should provide facilities to monitor the instances.
\item \textbf {BR22}: The system should provide facilities for logging the application's performance metrics for business-level troubleshooting.
\item \textbf {BR23}: The system should allow collaboration between different users to participate in the same project.
\item \textbf {BR25}: The system should avoid vendor lock-in.
\item \textbf {BR26}: The system should use computing resources efficiently to meet sustainability.
\item \textbf {BR27}: The provider should be able to support state-of-the-art hardware components for running the applications.
\item \textbf {BR28}: The provider should be able to support state-of-the-art software specifications to meet the requirements.
\item \textbf {BR29}: The system should be able to monitor and track the usage and allocation of resources.
\item \textbf {BR30}: The system should have a well-documented collection of the configuration instructions to allow seamless modifications.
\item \textbf {BR31}: The documentation should be available in as many languages as possible for global consumption.
\end{enumerate}

% ##########################################

\subsection{Technical Requirements} 
\label{Technical Requirements} 

Some of the important technical requirements that are considered to address the business requirements are detailed below. These requirements satisfy various pillars of the cloud architecture principles and broadly cover the whole functionality of the system.

\begin{enumerate}
\item \textbf{TR1 (BR1)}: The system should be able to read data that scales up to 100GB per day. 
\item \textbf{TR2 (BR2)}: The system should be able to store a minimum of 100TB of growing data in primary storage.
\item \textbf{TR3 (BR2)}: The system should have weekly archiving of data upon reaching 100GB into data warehouses.
\item \textbf{TR4 (BR3)}: The system’s messaging queues should be able to read incoming data of 1 GB per second and pass them to the storage and processing services.
\item \textbf{TR5 (BR4)}: The system should be scalable to accept and store data in textual and tabular form.
\item \textbf{TR6 (BR4)}: The system should be scalable to accept and store data in image and multimedia form.
\item \textbf{TR7 (BR4)}: The system should be scalable to accept and store data in file formats.
\item \textbf{TR8 (BR5)}: The storage provider should have the option of choosing the data storage from the top 100 countries with home country priority for sensitive and secure data.
\item \textbf{TR9 (BR5)}: The system should provide multiple physically separated and isolated Availability Zones, which are connected with low-latency, high-throughput, and highly redundant networking. It is important to distribute workload data and resources across multiple Availability Zones because the fault isolated boundaries limit the effect of a failure within a workload to a limited number of components. Components outside of the boundary are unaffected by the failure. Hence this TR directly relates to BR13 of having concrete data availability. \textbf{[1] (REL10 - Page 229)}

\item \textbf {TR10 (BR6)}: The cloud service provider should have a ‘pay as you use’ model that provides flexibility for the consumers. This directly relates to the BR6 for reducing the overall cost of the cloud system because the ‘pay as you use’ model allows to pay only for the computing resources that users require and increase or decrease usage depending on business requirements, not by using elaborate forecasting. For example, if the resources are only used for eight hours a day during the work week, then they can be stopped when they are not in use for a potential cost savings of 75\% (40 hours versus 168 hours). \textbf{[1] (Page 30)}.
\item \textbf {TR11 (BR6)}: The provider should allow the users to have an awareness of the ongoing and future expenditure. Policies should cover cost aspects of resources and workloads, including creation, modification and decommission over the resource lifetime. \textbf{[1] (COST02-BP01 Page 343)}
\item \textbf {TR12 (BR6)}: The system should enable the usage of appropriate services, resources and configurations for the workloads to optimize the overall cost spent on the services. 
\item \textbf {TR13 (BR7)}: The cost of maintaining a resource must be less than 100\$ per annum under the annual maintenance policy. 
\item \textbf{TR14 (BR8)}: The number of labor hours required to maintain a resource should not be greater than 100 hrs per annum.

\item \textbf{TR15 (BR9)}: The system should be available 99.999\% of the time with an average downtime of 3 minutes per month.  
\item \textbf{TR16 (BR9)}: The system should eliminate a single point of failure to ensure availability of 99.999\%.
\item \textbf{TR17 (BR9)}: The instances in the system should deploy load balancer to receive and route the requests to any available server at any point of time to ensure availability of 99.999\%. The reason why this technical requirement is related to the business requirement 9 is because the load balancer ensures the balanced distribution of traffic, thus enhancing the availability stability. \textbf{[1] (Page 317), [4]}
\item \textbf{TR18 (BR9)}: The system should be available 99.999\% of the time against physical events, such as natural disasters.
\item \textbf{TR19 (BR10)}: The system should have a 99.999\% throughput with a maximum acceptable latency under 3 seconds.
\item \textbf{TR20 (BR11)}: The response time of the system for requests reaching the core application should be under 10 milliseconds.

\item \textbf{TR21 (BR11)}: The system should query the database servers with a latency of not more than 100 milliseconds.
\item \textbf{TR22 (BR12)}: The system should ensure data restoration happening with a maximum accepted data loss of 1\% of the data generated within that day. 
\item \textbf{TR23 (BR12)}: The \textbf{Mean Time to Repair (MTTR)} for any major flaw of a component of the system should be on an average of 9 hours across any time of the year. This TR directly translates to having a highly available system by ensuring a quick recovery time because Workloads with a requirement for high availability and low mean time to recovery (MTTR) must be architected for resiliency. \textbf{[1] (Page 22)}
\item \textbf{TR24 (BR12)}: The \textbf{Mean Time between Failures (MTBF)} should be really high for any component of the system.
\item \textbf{TR25 (BR13)}: The provider should have three copies of data in two locations, one of which should be offsite. 
\item \textbf{TR26 (BR13)}: The service provider should have a data backup policy in place that calls for creating two copies of the original data—one for speedy recovery and the other to store in a safe, offsite location.

\item \textbf{TR27 (BR13)}: The system should be configured to take incremental backups of data every 12 hours. This TR directly relates to BR13 because taking regular backup of the data and testing the backup files ensures recovery from both logical and physical errors. \textbf{ [1] (Page23)}
\item \textbf{TR28 (BR13)}: The system should ensure that the backed up data is encrypted when in flight and when at rest.
\item \textbf{TR29 (BR14)}: The system should support CI/CD processes to have iterative and rapid deployment. Adopting approaches that improve the flow of changes into production enable refactoring, fast
feedback on quality, and bug fixing. \textbf{[1] (Page 9)}

\item \textbf {TR30 (BR14)}: The system should perform deployment only during scheduled release windows to prevent unnecessary downtime and induction of bugs.
\item \textbf {TR31 (BR14)}: The system should allow usage of scripts for the operations procedures and automate their execution by triggering them in response to events. These accelerate beneficial changes entering production, limit issues deployed, and enable rapid identification and remediation of issues introduced through deployment activities or discovered in your environments. \textbf{[1] (Page 9)}
\item \textbf {TR32 (BR14)}: The system should ensure that all the servers have the exact version of the code base to prevent some requests being served differently from others.
\item \textbf {TR33 (BR15)}: The system should be built on top of auto scaling tools to support automatic scaling of cloud resources based on the load. This directly relates to BR15 about having systems scale dynamically because Auto Scaling provides recommendations for scaling strategies customized to each resource. After creating a scaling plan, it combines dynamic scaling and predictive scaling methods together to support your scaling strategy. \textbf{[6] (Page 3)}
\item \textbf {TR34 (BR15)}: The system should scale out by one instance if average CPU usage is above 70\%, and scale in by one instance if CPU usage falls below 50\%. Specifying a scaling strategy helps to optimize the performance of the application based on the metric. This option helps you maintain high availability while reducing costs, ensuring resource availability during unpredictable scaling and handle reduced buffer capacities. \textbf{[6] (Page 10)}
\item \textbf {TR35 (BR15)}: The system should scale out to 10 instances on weekdays, and scale in to 4 instances on Saturday and Sunday.

\item \textbf {TR36 (BR16)}: The system shall provide identity management functionality to distinguish the input requests from different tenants. The reason why this technical requirement is related to the BR16 is because the technical requirements provide a requirement guideline to satisfy the business requirement on “identify and distinguish the requests from each user (tenant) uniquely”. \textbf{[1] (Page 14)}
\item \textbf {TR37 (BR17)}: The system should have a public key infrastructure to replace passwords through key management configuration. The reason why the technical requirement is related to the BR17 is because the encryption ensures the confidentiality of the information within the system that is hosted in the cloud environment. \textbf{[5] (Azure Storage encryption for data at rest)}
\item \textbf {TR38 (BR18)}: The system shall use encryption mechanisms to protect all the data residing on cloud storage.

\item \textbf {TR39 (BR18)}: The access to the data should be role restricted. The TR directly connects to BR 18 because the usage of such mechanisms and tools can reduce or eliminate the need for direct access or manual processing of data. This reduces the risk of mishandling or modification and human error when handling sensitive data. [1] (Page 13).
\item \textbf {TR40 (BR18)}: The system should have the data storage classified into sensitivity levels to protect the data accordingly. The reason this TR directly connects to BR18 is that the data should be protected in any form; hence it is important to classify the data into sensitivity levels (here they could be read or write databases) and use mechanisms, such as encryption, tokenization, and access control where appropriate. [1] (Page 13)
\item \textbf {TR41 (BR18)}: The system should have a Data Loss Prevention policy to monitor leakage or misuse of data.
\item \textbf {TR42 (BR19)}: The system should ensure that the ROOT account is secured. This directly relates to BR19 because only if the ROOT access is secure does it prevent direct penetration into the system. Only use the root user to perform tasks that specifically require it and periodically change the password of the ROOT. \textbf{[1] (Page 129)}
\item \textbf {TR43 (BR19)}: Enable some workloads in the system to be served from specific geographical locations to prevent attacks from hackers, worms, and external threats.
\item \textbf {TR44 (BR19)}: The system should be able to detect misconfigurations, vulnerabilities, and data security threats while providing actionable insights and guided remediation. This directly relates to BR19 to have a safe and secure system because along with defining the security strategies, it is important to have a mechanism that periodically takes action on security events and potential threats to help secure workload. \textbf{[1] (Page 137)}
\item \textbf{TR45 (BR19)}: Cloud Networking WAN should be separated into TWO Virtual Networking Environments, one for trusted users and another for untrust users.

\item \textbf{TR46 (BR20)}: Internal users and outside vendors should only be able to access the files needed to accomplish their work through the system.
\item \textbf{TR47 (BR20)}: The system should employ 2FA to prevent malicious login activities. The reason why this technical requirement helps achieve the goal of BR20 is that with multi-factor authentication, login activities can be secured and verified, thus preventing unwanted access. \textbf{[1] (Page 14)}
\item \textbf{TR48 (BR20)}: The system should enable firewall settings to restrict access to the infrastructure for untrust users.
\item \textbf{TR49 (BR20)}: Secure every endpoint (API; hence the applications running in the cloud systems. The reason why this technical requirement fits the BR20 is that the aspect provides encryption in transit when communicating with the API, thus ensuring the security of the endpoints of the application on the cloud architecture. \textbf{[1] Page 160 (SEC09-BP02 Enforce encryption in transit)}

\item \textbf {TR50 (BR21)}: The system should allow tracking of the access requests.
\item \textbf {TR51 (BR21)}: The system should be able to monitor virtual networks and virtual machine operations.
\item \textbf {TR52 (BR21)}: The system should provide dashboards to track processes, traffic, availability, and resource utilization. Designing the system in this way so that it provides the information necessary across all components (for example, metrics, logs, and traces) to understand its internal state. \textbf{[1] (Page 9)}
\item \textbf {TR53 (BR21)}: The system should include monitoring of the cloud costs to help make sure that scaling events doesn't cause users to cross the budget thresholds. The reason why the technical requirement fulfills the BR21 is that the monitor provides the ability for the cloud architect to have the infrastructure to monitor the budgeting and the service usage on the cloud infrastructure. \textbf{[1] (Page 32)}
\item \textbf {TR54 (BR21)}: The system should have a robust alerting system that provides all the information necessary  including stating what is monitored and its business impact for the on-call person to act immediately. The reason why this technical requirement fulfills the BR21 is because the alert in place enables the traceability of the system status and monitoring status. \textbf{[1] (Page 13)}
\item \textbf {TR55 (BR21)}: The system should be able to identify the person or team responsible for responding to the alert. Define in advance the personnel required to resolve an event and include escalation triggers to engage additional personnel, as it becomes necessary, based on urgency and impact. \textbf{[1] (Page 10)}

\item \textbf {TR56 (BR22)}: The system should always enable Admin Activity logs containing log entries for API calls or other administrative actions that modify the configuration or metadata of resources.
\item \textbf {TR57 (BR22)}: The system should have Data Access audit logs to record API calls that create, modify, or read user-provided data.
\item \textbf {TR58 (BR22)}: The system should allow monitoring the amount of requests being served and data being flown into the system.
\item \textbf {TR59 (BR23)}: The system should provide a version controlled platform for seamless collaboration of all the users.
\item \textbf {TR60 (BR24)}: Components on all layers of the system should ensure isolation between all components and functionalities. The reason why this technical requirement fulfills the BR24 is because, according to the reference, loose coupling helps isolate behavior of a component from other components that depend on it, increasing resiliency and agility. Failure in one component is isolated from others. With the isolated components, the reliability and resilience can be enhanced in the cloud architecture. \textbf{[1] (Page 190)}
\item \textbf {TR61 (BR24)}: The infrastructure should be secured through a VPC, having isolation of physical hosts and controlling network traffic.
\item \textbf {TR62 (BR25)}: The cloud provider must produce a clear exit strategy to prevent lock-in of data, resources and other sensitive information with the system.

\item \textbf {TR63 (BR25)}: The system should be cloud portable to enable organizations to migrate a cloud-deployed asset to a different provider to overcome vendor lock-in.
\item \textbf {TR64 (BR25)}: The provider should not create a non-competence agreement with the consumers.

\item \textbf {TR65 (BR26)}: The system’s carbon emissions need to be more effectively recorded and reduced. This TR directly connects to BR26 because it is important to establish sustainability goals while architecting the systems for the organization. Compare the productive output with the total impact of the cloud workloads by reviewing the resources and emissions required per unit of work. \textbf{[1] (Page 35)}
\item \textbf {TR66 (BR26)}: The system should power its operation by using renewable energy resources. As the technology keeps improving, it is very important to keep replacing the older resources with new energy powered ones. Continually monitor and evaluate new, more efficient hardware and software offerings. Design for flexibility to allow for the rapid adoption of new efficient technologies. \textbf{[1] (Page 35)}
\item \textbf {TR67 (BR27)}: The system should employ the latest version of Servers, Mobile devices, Hard drives, Network cables, Printers, Storage devices and Laptops.
\item \textbf{TR68 (BR28)}: The system should support both Windows \& Linux Distribution, SQLite and other state of the art software tools.
\item \textbf {TR69 (BR29)}: The system should tag every resource to easily identify and track the resources. Tagging the resources help to understand the organization, cost accounting, access controls, and targeting the execution of automated operations activities. \textbf{[1] (Page 10)}
\item \textbf {TR70 (BR29)}: The system should support and monitor over and under provisioning of resources when the demand increases and decreases respectively.
\item \textbf {TR71 (BR29)}: The system should effectively resolve the ties when two or more applications intend to use the same computing resources in the same instance.

\item \textbf {TR72 (BR30)}: The provider should clearly mention the steps to get started with using the systems and configure each component accordingly.
\item \textbf {TR73 (BR30)}: The document should contain code samples (design samples) to help users expedite the process of deploying their applications.
\item \textbf {TR74 (BR31)}: The document about the usage of the cloud services should be available in over 20 languages to be utilized by users across the globe.

\end{enumerate}

\subsection{Tradeoffs} 

Some of the important trade offs that have been made based on analyzing the technical requirements are:

\begin{enumerate}
    \item \textbf{Resource Utilization vs Reliability}: To maintain a really \textbf{reliable} system, there should be seamless usage and integration of the resources. Hence \textbf{resources could be over utilized}.
    \begin{enumerate}
        \item \textbf{TR66 (BR26)}: The system should power its operation by using renewable energy resources.
        \item \textbf{TR26 (BR13)}: The service provider should have a data backup policy in place that calls for creating two copies of the original data—one for speedy recovery and the other to store in a safe, offsite location.
    \end{enumerate}
    
    \item \textbf{Cost vs Availability}: Some of the application services can be made as \textbf{cheap} as possible which might cause unexpected performance mishaps and downtime thereby \textbf{NOT} being \textbf{available 24*7}.
    \begin{enumerate}
        \item \textbf{TR12 (BR6)}: The system should enable the usage of appropriate services, resources and configurations for the workloads to optimize the overall cost spent on the services.
        \item \textbf{TR15 (BR9)}: The system should be available 99.999\% of the time with an average downtime of 3 minutes per month.
    \end{enumerate}
    
    \item \textbf{Cost vs Security}: Having \textbf{low cost compute resources} and infrastructure might affect the \textbf{security} as they would not provide enough protection.
    \begin{enumerate}
        \item \textbf{TR66 (BR26)}: The system should power its operation by using renewable energy resources.
        \item \textbf{TR48 (BR20)}: The system should enable firewall settings to restrict access to the infrastructure for the untrust users.
    \end{enumerate}
    
    \item \textbf {Security vs Throughput}: The system must \textbf{ensure encryption and security} of data in rest and motion by \textbf{compromise on the high throughput} and performance rate.
    \begin{enumerate}
        \item \textbf{TR47 (BR20)}: The system should employ 2FA to prevent malicious login activities.
        \item \textbf{TR19 (BR10)}: The system should have a 99.999\% throughput with a maximum acceptable latency under 3 seconds.
    \end{enumerate}
    
    \item \textbf{Availability vs Sustainability}: The provider should have an immense amount of data being backed up in several availability zones not worrying about the resource consumption.
    \begin{enumerate}
        \item \textbf{TR25 (BR13)}: The provider should have three copies of data in two locations, one of which should be offsite.
        \item \textbf{TR65 (BR26)}: The system’s carbon emissions need to be more effectively recorded and reduced.
    \end{enumerate}
    
    \item \textbf{Elasticity vs Security}: Cloud services that \textbf{spin up and down} in an elastic fashion can impact existing security workflows and require them to be reimagined. Since elastic systems are ephemeral, incident response may be impacted, for example when a server experiencing a \textbf{security issue} spins down as demand wanes.
    \begin{enumerate}
        \item \textbf{TR34 (BR15)}: The system should scale out by one instance if average CPU usage is above 70\%, and scale in by one instance if CPU usage falls below 50\%. 
        \item \textbf{TR49 (BR20)}: Secure every endpoint (APIs) of the applications running in the cloud systems.
    \end{enumerate}
    
    \item \textbf{Availability vs Cost}:
    \begin{enumerate}
        \item \textbf{TR23 (BR12)}: The Mean Time to Repair (MTTR) for any major flaw of a component of the system should be on an average of 9 hours across any time of the year.
        \item \textbf{TR13 (BR7)}: The cost of maintaining a resource must be less than 100\$ per annum under the annual maintenance policy. 
    \end{enumerate}
    
    \item \textbf{Security vs Operational Excellence}:
    \begin{enumerate}
        \item \textbf{TR42 (BR19)}: The system should ensure that the ROOT account is secured.
        \item \textbf{TR31 (BR14)}: The system should allow usage of scripts for the operations procedures and automate their execution by triggering them in response to events.
    \end{enumerate}
 
\end{enumerate}

\newpage
\section{Provider Selection}
\label{sec:Provider Selection}

There are multiple Cloud Service Providers (CSPs) available out in the market who offer competitive services for addressing most of the generic requirements applicable to business applications. Some of the bigger names in the industry are \textbf{Amazon Web Services (AWS)}, \textbf{Google Cloud Platform (GCP)}, \textbf{Microsoft Azure}, \textbf{IBM Cloud, Oracle Cloud}, \textbf{Digital Ocean}, \textbf{CoreWeave} etc. However, choosing among such a plethora of alternatives is entirely based on the application's business and technological requirements. In Addition, since there is no designated framework for assessing the CSP, which increases the challenge for identifying the right provider.

This section will detail comparing the available providers on some of the important criteria and selecting a final provider to proceed with the rest of the architecture design. The providers enlisted are:
\begin{enumerate}
    \item Amazon Web Services - AWS
    \item Google Cloud Platform - GCP
    \item Microsoft Azure
    \item Coreweave
\end{enumerate}

%% YP: I am not following the last sentence "You must include at least three providers." Does this mean that they have investigate three providers and explain why they selected a provider?
%%YV: Yes; my initial thought was to force them to read the service descriptions/lists from the big three. I assumed three students per group, hence the three providers stipulation.

%% YP: How do we provide credits to the students? For example, if we plan to provide credits for AWS & GCP then will the student have to pay for Microsoft Cloud if they want to use that one? I believe we should somehow figure out how we can get the credits and provide them to the students. Then we can ask them to investigate the services between those two cloud providers. This work must be completed fairly early though because each cloud provide will have its own onboarding process and overhead for the instructors.
%%YV: I plan to get in touch with AWS for that. I know that in all likelihood, IBM will give the students infinite credit. If we have time during summer to contact GCP and Azure, we can certainly ask them. 
%%YV: for this step only, the students do not need any credits, I believe. They just check out the list of services. At the end, they can all use AWS, if this is the only provider who'll give us credits.

\subsection{Criteria for choosing a provider}
The providers are compared on a number of factors. Since most of the major providers offer the services in different versions suitable according to their operational model, it is important to evaluate these services and perform several trade offs between the services before deciding the provider. Some of the important criteria for choosing the provider for this business application is:

\begin{enumerate}
    \item \textbf{Security}
    
    Security is a top concern in the cloud and so it’s critical to ask detailed and explicit questions that relate to the unique business use cases, industry and regulatory requirements. Security should be inclusive of data, infrastructure, network and other intellectual properties of the application.

    \item \textbf{Data Storage}

    The provider should be able to support the storage and backup of extremely large volumes of data at seamless pace. The integrity and reliability of data should be highly prioritized.

    \item \textbf{Cost}

    There could be no dispute that pricing will be a significant consideration in determining which cloud service provider(s) to select. It's beneficial to consider both the asking price and associated expenses (including maintenance cost, one-time setup charges etc). For some of the components and features, it is advisable to choose low cost providers as they might just specialize with the particular functionality.

    \item \textbf{Performance}

    The services should have high throughput, availability and less latency.

    \item \textbf{Reliability}

    The services should be highly reliable and the provider must have a well planned mechanism of backup and recovery to ensure the data, application and infrastructure is always reliable within the system. 

    \item \textbf{Automation and Kubernetes}

    The provider should have modern and efficient automation and kubernetes setup that helps orchestrate the process of deployment, scaling, configuration management and service discovery.

    \item \textbf{Identity management}
    
    IAM (Identity and Access Management) is one of the most important yet complicated layers of cloud security. 

    \item \textbf{Availability Zones} 
    
    The cloud provider must be made available across multiple regions.
\end{enumerate}

\subsection{Provider Comparison} 
This section provides a detailed comparison for the different criterion listed above over all the major providers and ranks the providers in the order of the most likely preferred to the least preferred.

\begin{enumerate}
    \item \textbf{Security Comparison}

    Three factors that are considered the most important things when evaluating cloud vendors for security are: 
    \begin{enumerate}
        \item Physical security - Protecting data centers.
        \item Infrastructure security - Monitoring network traffic, and patching vulnerabilities. 
        \item Data and access controls - Controlling who has access to data; encryption.
    \end{enumerate}

    \begin{center}
    \begin{tabular}{|m{12em}|m{12em}|m{12em}|} 
      \hline
      \textbf{AWS} & \textbf{AZURE} & \textbf{GCP} \\ 
      \hline
      Most mature cloud provider. & The Azure infrastructure security is state-of-the-art and designed to manage extensive site-to-site connections hosting virtual networks across various regions. & Newness and the lack of experts trained on the platform as compared with AWS and Azure. \\ 
      \hline
      Network firewalls at layers 3, 4, and 7 built into Amazon Virtual Private Cloud (VPC) that allows customers to access individual instances and applications. & Security development lifecycle (SDL) -  reducing the number and severity of vulnerabilities in software, while reducing development cost. & Custom hardware and software in data centers and a strict hardware disposal policy. \\ 
      \hline
      Denial-of-service attacks (DoS) mitigation. & Intrusion and DoS detection. & Global IP network that minimizes the number of hops across the public internet. \\
      \hline
    
      Default encryption of all traffic between AWS facilities. & Network access control. & Security monitoring that is focused on internal network traffic. \\
      \hline
    \end{tabular}
    \end{center}

    All the 3 providers support most of the security requirements and provide similar functionalities. However Azure has the state of the art infrastructure where more customers can easily integrate with the Microsoft interface while GCP is pretty new and needs more training to be adept with the cloud. Hence the final ranking based on security is :
     $$ \textbf{AZURE} > \textbf{AWS} > \textbf{GCP} $$ 
    
    % ###############
    
    \item \textbf{Data Storage Comparison}
    
    \textbf{MLOPS} is heavily driven by the generation, consumption, and processing of data. Storing Big Data has become an essential component. Big data will be a combination of \textbf{THREE} different types of data - \textbf{structured}, \textbf{semistructured}, and unstructured. The characteristics of Big Data include:
    \begin{enumerate}
        \item The large \textbf{Volume} of data in many environments, such as transactions and financial records.
        \item A wide \textbf{Variety} of data types is frequently stored in big data systems, such as text, documents and multimedia files.
        \item The \textbf{Velocity} of the data is generated, collected, and processed, such as web server logs and streaming data from sensors.
    \end{enumerate}

    Big data is often stored in a Data Lake. \textbf{Amazon}, \textbf{Microsoft}, and \textbf{Google} all provide data lake technologies and solutions.

    \begin{center}
    \begin{tabular}{|m{9em}|m{9em}|m{9em}|m{9em}|} 
      \hline
      \textbf{Criteria} & \textbf{AWS} & \textbf{AZURE} & \textbf{GCP} \\ 
      \hline
      Solutions, Services and Tools & Data Lake on AWS [7], AWS Lake Formation, AWS S3. & HDInsight, Data Lake Analytics and Azure Data Lake Storage [8]. & Data Lake Modernization, BigLake [9], Dataflow, BiqQuert, Cloud Data Fusion and Dataproc.\\
      \hline
      Data Management & Available & Available & Available \\
      \hline
      Pay as you use & Available & Available & Available \\
      \hline
      USP & Support storage and analytics at one place. & Ease of Use, cheap and most promising. & Provide extended multi cloud support. \\
      \hline
    \end{tabular}
    \end{center}

    Upon comparing all the providers, from \textbf{the ease of use aspect}, \textbf{Azure Data Lake Gen2} and \textbf{GCP BigLake} win. BigLake could sit in BigQuery on AWS S3, and Azure Data Lake Storage Gen2 is the most promising one. However, BigLake is still in preview. 

    $$ \textbf{Azure Data Lake Gen2} >=  \textbf{GCP BigLake} >= \textbf{Data Lake on AWS} $$

    % ###########################
    
    \item \textbf{Cost Comparison}

    \begin{center}
        \begin{tabular}{|c|c|c|c|}
            \hline
            \textbf{Criteria} & \textbf{AWS} & \textbf{AZURE} & \textbf{GCP} \\ 
            \hline
            Storage (per GB/Month) & \$0.023 & \$0.021 & \$0.023 \\
            \hline
            General Purpose compute & \$0.1344 & \$0.166 & \$0.150924 \\
            \hline
            Optimized compute & \$0.153 & \$0.1690 & \$0.2351 \\
            \hline
            General purpose discounted & \$0.079 & \$0.0974 & \$0.095092 \\
            \hline
            Optimized compute discounted (per hour) & \$0.094	& \$0.10 & \$0.13156 \\
            \hline
            Spot Instances VM - Optimized Compute & \$0.068 & \$0.0259 & \$0.0540 \\
            \hline
            Accelerated Computing (per hour) & \$0.90 & \$0.526 & \$3.678 \\
            \hline Total price (per month) & \$18.74 & \$18 & \$25.15 \\
            \hline
        \end{tabular}
    \end{center}

    The cost for storage, optimized compute at discounted price, accelerated compute and spot instances VM are cheap for Microsoft Azure. Also the overall price per month is cheapest for Azure as compared to AWS and GCP. Hence the ranking for the general cost is given by
    $$ \textbf{AZURE} >= \textbf{AWS} > \textbf{GCP} $$
    
    \begin{center}
        \begin{tabular}{|c|c|c|c|c|c|}
        \hline
        \textbf{GPU Cloud} & \textbf{GPU} & \textbf{Marketing Name} & \textbf{GPU Memory} & \textbf{Tenser} & \textbf{Price(hour)} \\
        \hline 
        Coreweave & A100 40GB & A100 40GB & 40GB & 432 & \$2.06 \\
        \hline 
        Azure & A100 40GB & NC24ads A100 v4 & 40GB & 432 & \$3.67 \\
        \hline 
        AWS & A100 40GB & EC2 P4d & 40GB & 432 & \$4.10 \\
        \hline 
        Google & A100 40GB & & 40GB & 432 & \$3.67 \\
        \hline
        \end{tabular}
    \end{center}
    Specific to the application of MLOPS, there is a requirement to make use of a high processing GPU with respect to Memory and number of tensors. For this purpose, \textbf{Coreweave} (a provider specializing in GPUs for ML infrastructure) provides the infrastructure at the cheapest cost. 

    \item \textbf{Performance Comparison}
    
    \textbf{GCP}’s top-performing machine had 165\% and 237\% more \textbf{throughput} than AWS and Azure respectively. GCP simply made more bandwidth available (up to 32 Gbps) than AWS (up to 10 Gbps, or in some cases up to 25 Gbps) or Azure (up to 8 Gbps).
    
    \textbf{AWS} has performed best in network latency with its top-performing machine’s 99th percentile network latency was 28\% and 37\% lower than Azure and GCP, respectively.
    
    \textbf{Azure}’s ultra disk meanwhile delivered 16\% more TPM while being priced only 11\% higher than Azure’s less expensive “premium” disks.
    $$ \textbf{GCP} = \textbf{AWS} = \textbf{AZURE} $$

    \item \textbf{Reliability Comparison}

    \begin{center}
        \begin{tabular}{|m{9em}|m{9em}|m{9em}|m{9em}|}
        \hline
        \textbf{Criteria} & \textbf{AWS} & \textbf{AZURE} & \textbf{GCP} \\
        \hline 
        SLA Level & 99.995\% & 99.95\% (For 2+ VMs running in the same availability set) & 99.95\% \\
        \hline 
        SLA Credit & 10\% of annual bill & $\leq$ 99.9\% - 10\%, $\leq$ 99\% - 25\% & 10\% - 50\%  \\
        \hline 
        Multiple zones in region & 2-5 per region & Some geographic regions have multiple “regions” in Azure. Azure automatically allocates instances in the same role or availability set into different fault domains & Available. \\
        \hline 
        Automatic Failover to other server & Partial (Auto recovery services applies only to C3, C4, M3, R3 and T2 instances). & Available (Live migration allows moving instances to another server without downtime) & Available (Host maintenance does not affect cloud availability). \\
        \hline 
        Storage attachment & Instance storage: server, EBS: network & network & network \& local \\ 
        \hline
        Backup Snapshots & Available & Available & Available\\
        \hline
        Backup storage & Available (S3 or Glacier) & Available & Available \\
        \hline

        \end{tabular}
    \end{center}

    Azure has been equally well placed with AWS in terms of providing backup snapshots, storage and an almost equal SLA level of 99.99\%. However, it clearly beats AWS with the provision of automatic failover to other servers which is a key aspect of Reliability. Hence, the ranking would be:
    $$ \textbf{AZURE} = \textbf{AWS} >= \textbf{GCP} $$
    
    \item \textbf{Kubernetes \& Automation Comparison}
    
    Kubernetes will manage all cloud sub-systems.

    \begin{center}
        \begin{longtable}{|m{9em}|m{9em}|m{9em}|m{9em}|}
        \hline
        \textbf{Service} & \textbf{AWS} & \textbf{AZURE} & \textbf{GCP} \\
        \hline
        Current Version (Updates) & Least frequent releases. & Integrated with Microsoft hence receives 2nd most frequent releases. & Original creator of Kubernetes, so introduction of new features is faster. \\
        \hline
        Automatic update & Upgrading Amazon EKS, a user needs to send some command-line instructions to it, which makes it more difficult among the other two. & AKS allows upgrading the cluster by a simple command. & GKE is at the top as it provides a fully automated update for the cluster. \\ 
        \hline
        Resource Monitoring & Cloud watch. & Microsoft Azure has two offerings: Azure Monitor to evaluate the health of a container and Application Insights to monitor the Kubernetes components.  & Google Cloud provides Stackdriver. Stackdriver monitors the master and nodes, and all Kubernetes components inside the platform along with integrating logging. \\
        \hline
        Autoscaling & AWS is ranked as second in auto-scaling because it needs some minor manual configurations.  & Microsoft Azure has introduced autoscaler, which is partially covered by customer support  & Google Cloud is leading as the most mature solution available on the interface. \\
        \hline
        Cost & \$.20 per hour & Only pay for the VMs running the Kubernetes nodes. & Only pay for the VMs running the Kubernetes nodes.\\
        \hline
        Ease of usage & Difficult to use for new developers unfamiliar with container services and Kubernetes. & More intuitive for Windows developers, Supports Windows and Linux containers & Doesn’t integrate with IaaS cloud requirements. Works well with DevOps teams.\\
        \hline
        High-Availability Clusters & No & In development & Yes \\
        \hline
        \end{longtable}
    \end{center}

    Although most of the service providers offer equally comparable Kubernetes services, Azure is preferred here because it has a very accommodative policy for the pricing of  VMs running on Kubernetes nodes. Also usage of Azure would be very intuitive supporting Windows based development.
    $$ \textbf{AWS} >= \textbf{GCP} >= \textbf{AZURE} $$
    
    \item \textbf{Identity Management Comparison}

    \begin{center}
        \begin{longtable}{|m{9em}|m{9em}|m{9em}|m{9em}|}
        \hline
        Criteria & AWS & AZURE & GCP \\ 
        \hline
        Resource Structure & Main container is called an AWS account, which can be set up and used to provision resources. & Azure (and GCP) employ an RBAC (role-based access control) model, which relies on a more methodical structure of resources. Azure provides a basic resource container, the “resource group.”  & In GCP, the basic container is the “project,” which contains the resources for a single application. A “folder” is a container that can aggregate several projects.  \\ 
        \hline
        Listing Permissions & In AWS, the permissions document, named “IAM Policy,” includes the details of the permissions and the resources in the same JSON file.  & The list of permissions is decoupled from the resources. The resources for which a permission assignment would apply is called a "scope."  & GCP also separates the permissions from the resources in the document named “role.” The role document does not list resources. \\ 
        \hline
        User Access Management & In AWS, admins can create a resource named “IAM user,” an object that represents the resource and an IAM policy that determines access capabilities. & Azure offers the Azure Active Directory (Azure AD) – the identity provider where users are managed. This includes creating new users, inviting guest users or managing how they access applications. & For IAM in GCP, users are managed through Google Cloud Identity or Google Workspace. In addition, GCP has additional entities that can be given permissions, including domain, service accounts, authenticated users and all users. \\
        \hline
        Service Access Management & In AWS, service identities can be given access to resources by assumption of IAM roles. & Azure provides two options for managing service access. The first is for giving resource access to services by creating a “managed identity,” which enables granting access without needing to manage credentials. The second option for providing access in Azure is by using a “service principal.” & GCP enables providing access to a type of proxy identity called a “service account”. \\
        \hline
        Pitfalls & Static Credentials in AWS. Security issues & In Azure, multiple resources have static credentials, called “access keys”. & Static Credentials in GCP. \\
        \hline
        Security Issues. & AWS uses a multi-stage, complex process when evaluating whether to give permissions. & Azure provides multiple guardrails however most are fairly straightforward. \textbf{(Locks, Deny Assignments, Conditional Access, Azure Policy)} & An effective GCP guardrail is the IAM Deny policy. This policy is a set of rules that determines what a principal is denied access to. \\
        \hline
        \end{longtable}
    \end{center}
    Azure is preferred over AWS and GCP because it has a two level strategy for service level management and also has a fairly straightforward implementation of the guardrails which can set up the identity management process seamlessly. In terms of other functionalities, all the providers are similar to each other.
    $$ \textbf{AZURE} >= \textbf{AWS} = \textbf{GCP} $$
    
    \item \textbf{Availability Zones Comparison}

    \begin{center}
        \begin{tabular}{|c|c|c|}
            \hline
            \textbf{AWS} & \textbf{AZURE} & \textbf{GCP} \\
            \hline
            66 zones & 54 regions worldwide in 140 countries & 20 regions \\
            \hline
        \end{tabular}        
    \end{center}
    AWS has more coverage and they are rapidly expanding the availability zones. However, Azure is not fairly far behind as they have zones over 140 countries.
    
    $$ \textbf{AWS} >= \textbf{AZURE} > \textbf{GCP} $$

\end{enumerate}

\subsection{The final selection} 
Based on the different criteria mentioned in the previous section and the specificalities of the Business and Technical Requirements, there is no one clear winner that emerged. Some of the criteria like Availability Zones and Kubernetes have AWS providing extremely good services. Other services like Reliability, Performance, Storage and Security have equally beneficial services provided by all the major providers. However, having the idea of Cost and Identity Management being prioritized, we proceed forward with a \textbf{MICROSOFT AZURE} as our \textbf{primary cloud service provider}. The training Cluster will be designed using \textbf{Coreweave} provider.

\subsubsection{The list of services offered by the winner} 
Microsoft Azure offers an array of services that would be helpful to implement various requirements related to Security, Performance, Database storage and Kubernetes. Some of the services are listed in detail below:

\textbf{Azure Firewall} is a cloud-native and intelligent network firewall security service that provides the best of breed threat protection for cloud workloads running in Azure. It's a fully stateful, firewall as a service with built-in high availability and unrestricted cloud scalability. It provides both east-west and north-south traffic inspection. {\textbf{[18]}}

\textbf{Azure VPN Gateway} sends encrypted traffic between an Azure virtual network and an on-premises location over the public Internet. The VPN Gateway can also be used to send encrypted traffic between Azure virtual networks over the Microsoft network. When creating multiple connections to the same VPN gateway, all VPN tunnels share the available gateway bandwidth. {\textbf{[21]}}

\textbf{Azure VNet Hub and Spoke} virtual network acts as a central point of connectivity to many spoke virtual networks. The hub can also be used as the connectivity point to the on-premises networks. The spoke virtual networks peer with the hub and can be used to isolate workloads. {\textbf{[16]}}

\textbf{Virtual networking} is the abstraction of network resources from the underlying physical network infrastructure. A virtual network is the logical representation of the physical hardware, such as the switches and routers, that makes up the infrastructure. {\textbf{[16]}}

\textbf{Azure Monitor} helps maximize the availability and performance of the applications and services. It delivers a comprehensive solution for collecting, analyzing, and acting on telemetry from cloud and on-premises environments. This information helps administrators understand how the applications are performing and proactively identify issues that affect them and the resources they depend on. {\textbf{[23]}}

\textbf{Azure Kubernetes Service (AKS)} simplifies deploying a managed Kubernetes cluster in Azure by offloading the operational overhead to Azure. As a hosted Kubernetes service, Azure handles critical tasks, like health monitoring and maintenance. Since Kubernetes masters are managed by Azure, the cloud consumer only manages and maintains the agent nodes. Thus, AKS is free; cloud consumers only pay for the agent nodes within the clusters, not for the masters. {\textbf{[26]}}

\textbf{Azure Active Directory (Azure AD)} is a cloud-based identity and access management service. Azure Active Directory also helps them access internal resources like apps on corporate intranet networks, along with any cloud apps developed for organization. {\textbf{[13]}

\textbf{Azure Data Lake Storage Gen2} is a set of capabilities dedicated to big data analytics, built on Azure Blob Storage. {\textbf{[20]}

\textbf{Azure Automation} delivers a cloud-based automation, operating system updates, and configuration service that supports consistent management across Azure and non-Azure environments. It includes process automation, configuration management, update management, shared capabilities, and heterogeneous features. {\textbf{[14]}
\newpage

\section{The first design draft}
\label{sec:first design draft}

%% YP: It would be great for them to also put in the first draft a list of action items to complete the project. Potentially also include a rough timeline on how long each action item will take for them to complete it.   
%% YV: added.

Below is a list of the final set of technical requirements that were hand-selected from the larger range of requirements. These specifications address all of the design tenets of well-architected principles. Some of them are in conflict with one another, simulating situations in the actual world where trade-offs must be made when designing the system.

\textbf{\underline {Identity Management}}

\textbf{TR36 (BR16)} - The system shall provide identity management functionality to distinguish the input requests from different tenants

\textbf{\underline {Operational Excellence Pillar}}

\textbf{TR31 (BR14)} -  The system should allow usage of scripts for the operations procedures and automate their execution by triggering them in response to events.

\textbf{\underline {Security Pillar}}

\textbf{TR61 (BR24)} - The infrastructure should be secured through a VPC, having isolation of physical hosts and controlling network traffic. 

\textbf{TR48 (BR20)} - The system should enable firewall settings to restrict access to the infrastructure for the untrust users.

\textbf{TR45 (BR19)} - Cloud Networking WAN should be separated as TWO Virtual Networking Environments, one for trusted users and another for untrust users.

\textbf{\underline{Reliability Pillar}}

\textbf{TR28 (BR13)} - The system should ensure that the backed up data is encrypted when in flight and when at rest.

\textbf{\underline{Performance Efficiency Pillar (Includes TR for Monitoring)}}

\textbf{TR19 (BR10)}- The system should have a 99.999\% throughput with a maximum acceptable latency under 3 seconds.

\textbf{TR58 (BR22)} - The system should allow monitoring the amount of requests being served and data being flown into the system.

\textbf{\underline{Cost Optimization Pillar}}

\textbf{TR12 (BR6)}: The system should enable the usage of appropriate services, resources and configurations for the workloads to optimize the overall cost spent on the services. 

\textbf{\underline{Sustainability Pillar}}

\textbf{TR34 (BR15)} - The system should scale out by one instance if average CPU usage is above 70\%, and scale in by one instance if CPU usage falls below 50\%.

\textbf{\underline{Conflicting Pairs of Technical Requirements}}
\begin{enumerate}
    \item \textbf{Performance Efficiency vs Cost Optimization} - Monitoring is a crucial part of evaluating the system's performance, but it uses more resources and is more expensive than necessary.
    \begin{enumerate}
        \item \textbf{TR58 (BR22)} - The system should allow monitoring the amount of requests being served and data being flown into the system.
        \item \textbf{TR12 (BR6)}: The system should enable the usage of appropriate services, resources and configurations for the workloads to optimize the overall cost spent on the services. 
    \end{enumerate}

    \item \textbf{Performance Efficiency vs Cost Optimization} - All services, including databases, virtual machines, process engines, etc., must have load balancers deployed, which increases the cost.
    \begin{enumerate}
        \item \textbf{TR19 (BR10)}: The system should have a 99.999\% throughput with a maximum acceptable latency under 3 seconds.
        \item \textbf{TR12 (BR6)}: The system should enable the usage of appropriate services, resources and configurations for the workloads to optimize the overall cost spent on the services. 
    \end{enumerate}
    
    \item \textbf{Reliability vs Performance Efficiency} - If the data is encrypted, the throughput of the system gets affected, falling short of 99.999\% availability.
    \begin{enumerate}
        \item \textbf{TR28 (BR13)} - The system should ensure that the backed up data is encrypted when in flight and when at rest.
        \item \textbf{TR19 (BR10)}: The system should have a 99.999\% throughput with a maximum acceptable latency under 3 seconds.
    \end{enumerate}

    \item \textbf{Operational Excellence vs Security} - Multiple services on various hosts must be accessed for automation. As a result of their VPC security, their recurring requests may be controlled as part of the network traffic.
    \begin{enumerate}
        \item \textbf{TR31 (BR14)} -  The system should allow usage of scripts for the operations procedures and automate their execution by triggering them in response to events.

        \item \textbf{TR61 (BR24)} - The infrastructure should be secured through a VPC, having isolation of physical hosts and controlling network traffic.
    \end{enumerate}
    
\end{enumerate}

% ###########################
\subsection{The basic building blocks of the design} 
%Putting the blocks together
We divide all the services that need to be used into \textbf{FIVE} categories.
\begin{enumerate}
    \item Infrastructure
    \item Networking Services
    \item Container Related
    \item Storage
    \item Security Related
    \item Monitoring Related
\end{enumerate}

The Infrastructure contains \textbf{FOUR} services.
\begin{enumerate}
    \item Virtual Networking
    \item Virtual Subnet
    \item Spoke-Hub
    \item vWAN Hub
\end{enumerate}

The Networking Services contain \textbf{FIVE} services.
\begin{enumerate}
    \item Front Doors
    \item Application Gateway
    \item Firewall
    \item DNS
    \item Private Endpoint
\end{enumerate}

The Container Related contains \textbf{FOUR} services.
\begin{enumerate}
    \item Container Instances
    \item K8S
    \item Data Science VM
    \item Container Registries
\end{enumerate}

The storage contains \textbf{ONE} service.
\begin{enumerate}
    \item Azure Data Lake
\end{enumerate}

The Security and Tenant Identification Related contains \textbf{ONE} service.
\begin{enumerate}
    \item Active Directory
\end{enumerate}

The Monitoring related contains \textbf{THREE} service.
\begin{enumerate}
    \item Azure Monitoring
    \item Azure Alerts
    \item Azure Log Analytics
\end{enumerate}

% ###########################

\subsection{Top-level, informal validation of the design} 

\textbf{TR36 (BR16) - The system shall provide identity management functionality to distinguish the input requests from different tenants.}

The system should be able to recognize each tenant separately in order to manage their requests, the resources allotted to them, and the billing of users depending on usage. It is necessary to administer the Virtual Subnets for identity authentication. Microsoft Azure offers a service referred known as the \textbf{Azure Active Directory} [13] to meet this requirement. Any authentication solution supported by Azure AD, including \textbf{Key Vault}, can be used to authenticate Azure services [15].

\textbf{TR31 (BR14) -  The system should allow usage of scripts for the operations procedures and automate their execution by triggering them in response to events.}

Process automation in \textbf{Azure Automation} enables the automation of routine, labor-intensive, and error-prone management operations. This service aids in lowering operational costs by minimizing errors and concentrate on tasks that improve the bottom line. The service allows authors to write runbooks in graphical, PowerShell, and Python. Webhooks enable request fulfillment and provide continuous delivery and operations by triggering automation from Azure Logic Apps, \textbf{Azure Function}, ITSM goods or services, DevOps, and monitoring systems [14].

\textbf{TR61 (BR24) - The infrastructure should be secured through a VPC, having isolation of physical hosts and controlling network traffic.}

The \textbf{Azure Virtual Network} Service, which functions similarly to a VPC, makes it possible to protect host isolation and manage network traffic [16]. Using virtual network peering, this can link up two virtual networks, allowing their resources to talk with one another across regions. \textbf{Azure VPN Gateway} or \textbf{ExpressRoute} connections, used to connect the virtual network to an on-premises network, aid in spreading the on-premises BGP routes to the specified virtual networks [17]. 

\textbf{TR48 (BR20) - The system should enable firewall settings to restrict access to the infrastructure for the untrust users.}

\textbf{Azure Firewall} Standard offers threat intelligence feeds and L3-L7 filtering [18]. Threat intelligence-based filtering has the ability to warn about and block traffic from/to known malicious IP addresses. It is crucial to safeguard those connections as well because the program is accessible to third parties who might log on to the system via the web. Both \textbf{VNet} and \textbf{Virtual WANs} (Secure Virtual Hub) environments are supported by Firewall Manager for firewalls. 

\textbf{TR45 (BR19) - Cloud Networking WAN should be separated as TWO Virtual Networking Environments, one for Trusted users and another for Untrust users.}

\textbf{Azure Front Door} provides intelligent security that embraces a \textbf{Zero Trust framework}, enabling internet-facing applications to intelligently defend the digital estate against known and novel threats [19]. This enables them to distinguish between system users who are \textbf{Trusted} and those who are \textbf{Untrusted}. The on-premise network's data gathering subsystem is connected to the trust users. The customers from outside the system who attempt to access it via the open internet are the untrustworthy users.

\textbf{TR28 (BR13) - The system should ensure that the backed up data is encrypted when in flight and at rest.}

\textbf{The Data Lake Gen2} security paradigm provides POSIX permissions, ACLs, and more granularity tailored to Data Lake Storage Gen2. By utilizing Azure role-based access management, the \textbf{Azure Data Warehouse} gives users the ability to govern activities with fine-grained access (Azure RBAC). Between Azure storage and the vault, data in transit is secured via HTTPS within Azure [20].

\textbf{TR19 (BR10): The system should have a 99.999\% throughput with a maximum acceptable latency under 3 seconds.}

\textbf{API Gateway} serves as a reverse proxy, forwarding client requests to services [21]. Inbound flows are distributed to back-end pool instances by the \textbf{Azure Load Balancer} that flows follow the load-balancing rules that have been set up. All TCP and UDP applications may scale up to millions of flows with the help of a load balancer, which offers low latency and high throughput. Key scenarios that can be accomplished using Azure Standard Load Balancer [22] include:
\begin{enumerate}
    \item Load balance internal and external traffic to Azure virtual machines.
    \item Increase availability by distributing resources within and across zones.
    \item Use health probes to monitor load-balanced resources.
    \item Employ port forwarding to access virtual machines in a virtual network by public IP address and port.
\end{enumerate}

\textbf{TR58 (BR22) - The system should allow monitoring the amount of requests being served and data being flown into the system.}

\textbf{Microsoft Azure Monitor}, which stores metrics, logs, traces, and changes data, is used to understand the system's performance and bottlenecks and monitor the user interactions with the system [23]. This data is stored collectively, allowing for easy correlation and analysis with a shared set of tools. 

\textbf{TR12 (BR6): The system should enable the usage of appropriate services, resources and configurations for the workloads to optimize the overall cost spent on the services.} 

The \textbf{Azure Savings Plan} offers a customizable approach to get up to a 65\% savings on computing services [24]. The \textbf{Azure Hybrid Benefit} is a license perk that enables to accomplish more with less [25]. The first step in enterprise cost optimization is for the team to agree on the tools, procedures, and dependencies needed to respond wisely to cost issues at the organizational level. These outputs are the outcome of the following recurrent tasks:
\begin{enumerate}
    \item Ensure strategic alignment with the cloud strategy team (which includes workload stakeholders across the portfolio).
    \item Align spending to budget expectations.
\end{enumerate}

\textbf{TR34 (BR15) - The system should scale out by one instance if average CPU usage is above 70\%, and scale in by one instance if CPU usage falls below 50\%.}

The deployment of a managed Kubernetes cluster is made simpler by \textbf{Azure Kubernetes Service (AKS)}. AKS allows Kubernetes clusters that run various node pools. The amount of cluster nodes or pods that run the services dynamically adjusts up or down as the demand for resources varies. Cluster and horizontal pod autoscalers can both be modified to respond to demand and only use required resources [26].

\newpage

\section{The second design}
\label{sec:second design}

\subsection{Discussion of pillars} 

\subsubsection{Security}
The security pillar, according to the AWS WAF, explains how to use cloud technologies to protect the data, systems, and assets in a way that can enhance the security posture of the cloud solution. There are a number of fundamental design principles that deal with workload security that are contained inside the security framework. \textbf{[1] Page 14}

The security pillars include a wide range of subjects, including infrastructure protection, data protection, identity and access management, detection, and incident response. \textbf{[1] Page 13}

Eight best practices were suggested to solve the questions focused on the security foundations components. Security foundation practices from the AWS WAF include separating workloads with accounts, securing an AWS account, identifying and validating control objectives, staying up to date on security threats and recommendations, automating testing and validation of security controls in pipelines, identifying and prioritizing risks using a threat model, and routinely evaluating and implementing new security services and features. \textbf{[1] Page 117}

Best practices for managing authentication included using robust sign-in techniques, temporary credentials, securely storing and using secrets, relying on a centralized identity provider, auditing and rotating credentials on a regular basis, and utilizing user groups and attributes. In contrast to managing authentication, managing permissions places more of an emphasis on aspects of control from the perspective of the administrator than the system user. The AWS WAF recommended best practices for permission control, including defining access requirements, granting least privilege access, establishing emergency access procedures, reducing permissions continuously, defining permission guardrails for organizations, managing access based on lifecycle, analyzing public and cross-account access, and securely sharing resources. \textbf{[1] Page 145}

For the purpose of facilitating the detection and investigation of security events, best practices such as configuring service and application logging, centrally analyzing logs, results, and metrics, automating response to events, and implementing actionable security events are advocated. \textbf{[1] Page 29}

The focus of network resource protections and compute resource protections was central to all best practices for infrastructure protection. The creation of network layers, traffic management at all layers, automation of network protection, and implementation of inspection and protection were established best practices for the protection of network resources. Common best practices for protecting computing resources included vulnerability monitoring, lowering attack surfaces, implementing managed services, automating compute protection, enabling remote action, and validating software integrity. \textbf{[1] Page 147}

Data classification and protection both at rest and in transit are the main areas of attention for data protection. The most frequently recommended best practices for classifying data include defining data security measures, automating identification and categorization, and defining data lifecycle management. The AWS WAF also recommends practices to protect data in transit, including implementing secure key and certificate management, enforcing encryption in transit, automating detection of unauthorized data access, and authenticating network communications. \textbf{[1] Page 155}

An important part of incident responses is anticipating, responding to, and recovering from incidents. Common practices mentioned include identifying key personnel and outside resources, developing incident management plans, preparing forensic capabilities, automating containment capability, pre-provisioning access, pre-deploying tools, and running simulations on internal events that give a structured opportunity to practice incident management plans and procedures in a realistic scenario. \textbf{[1] Page 163}

Overall, to solve the majority of the security issues, AWS WAF offered a complete solution.

\subsubsection{Performance Efficiency}
The performance efficiency pillar, according to the AWS WAF, represents the capacity to employ computational resources effectively to satisfy system needs and to retain effectiveness as demand changes and technologies advance. The performance efficiency pillar generally covers a wide range of subjects, including monitoring, tradeoffs, review, and selection. \textbf{[1] Page 23}

Seven best practices were proposed to address the issues in the selection process, including understanding the services and resources available, defining a process for architectural choices, accounting for cost requirements in decisions, using policies or reference architectures, using guidance from cloud providers or appropriate partners, benchmarking existing workloads, and load testing workload. \textbf{[1] Page 283}

The AWS WAF document provided best practices for compute solution considerations, such as evaluating compute options, comprehending compute configuration options, collecting compute-related metrics, determining the required configuration by right-sizing, utilizing the available elasticity of resources, and re-evaluating compute needs based on metrics. \textbf{[1] Page 25, 294}

The recommended best practices for storage solutions include understanding storage features and requirements, assessing configuration alternatives, and basing decisions on access patterns and metrics. Best practices were recommended for understanding data characteristics, evaluating the possibilities, gathering and recording database performance measurements, selecting data storage based on access patterns, and optimizing data storage based on access patterns and metrics. \textbf{[1] Page 26}

The best practices for configuring networking solutions include understanding how networking affects performance, evaluating networking features, choosing a dedicated connection or VPN that is appropriately sized for hybrid workloads, utilizing load balancing and encryption offloading, choosing network protocols to improve performance, deciding where to locate a workload based on network requirements, and optimizing network configuration based on metrics. \textbf{[1] Page 27}

To solve the issue of monitoring resources to assure the intended performance, a number of best practices such as recording performance-related metrics, analyzing metrics when events or incidents occur, establishing key performance indicators (KPIs) to measure workload performance, using monitoring to generate alarm-based notifications, reviewing metrics on a regular basis, and monitoring and alarming proactively were mentioned. \textbf{[1] Page 28}

The best practices for facilitating the tradeoffs consideration process in the performance efficiency analysis include understanding the areas where performance is most crucial, learning about design patterns and services, identifying how tradeoffs impact customers and efficiency, measuring the impact of performance improvements, and using various performance-related strategies. \textbf{[1] Page 29}

Overall, the performance efficiency portion of AWS WAF offered a thorough overview of those key factors and offered analytical perspectives to make it easier to take performance efficiency into account when designing a cloud computing system.

\subsubsection{Cost Optimization}

According to the AWS WAF, the ability to operate systems to provide business value at the lowest price point is described by the cost optimization pillar. There are five best practices for cost optimization in the cloud, including cloud financial management, awareness of spending and consumption, cost-effective resources, demand and supply management, and long-term optimization. \textbf{[1] Page 31}

As the solution scales on AWS, enterprises are able to generate business value and financial success by deploying cloud financial management. The ability to link resource prices to specific companies or product owners encourages resource-efficient behavior and lowers waste. \textbf{[1] Page 31}

In order to be aware of expenditures and usage, it is common practice to implement change control and resource management from the start of a project to its end of life. These practices include creating policies and mechanisms to make sure that the right costs are incurred while the objectives are met. \textbf{[1] Page 32}

Best practices for cost-effective resources include making sure that the right size and number of resources are used for the task at hand, using the pricing model that is best for resources to minimize costs, and making sure to plan and monitor data transfer costs so that architectural decisions can be made to cut costs. \textbf{[1] Page 33}

Overall, the AWS WAF's cost optimization section offered an intriguing viewpoint and theoretical guidelines for cost reduction for the proposed cloud computing architecture.

\subsection{Use of Cloudformation diagrams} 

\begin{figure}[htp]
    \centering
    \includegraphics[width=15cm]{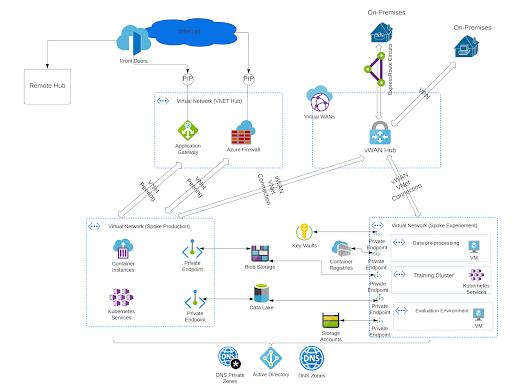}
    \caption{Cloud architecture diagram hosting the MLOPS application}
    \label{fig:galaxy}
\end{figure}

% #######################

\subsection{Validation of the design}

\textbf{TR36 (BR16) - The system shall provide identity management functionality to distinguish the input requests from different tenants.
}

\begin{figure}[htp]
    \centering
    \includegraphics[width=15cm]{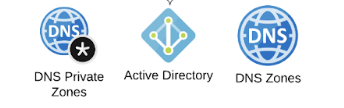}
    \caption{Identity Management}
    \label{fig:galaxy}
\end{figure}

Identity management is the process of authenticating and authorizing security principals. Users would be added to a resource group whenever they started using the Azure Portal to access their cloud systems. Each resource group has its own individual identification code, or \textbf{tenant id}, which is owned by the organization. A tenant id may be associated with more than one \textbf{subscription id}, and usage is recorded for each one. Therefore, by uniquely identifying each tenant and the requests, the design of these features utilizing \textbf{Azure Active Directory} and building resource groups for each organization assists in verifying the business and technical requirements.

By taking advantage of the security benefits of Azure Active Directory [13] (Azure AD), one can:
\begin{enumerate}
    \item Create and manage a single identity for each user across your hybrid enterprise, keeping users, groups, and devices in sync.
    \item Provide SSO access to your applications, including thousands of pre-integrated SaaS apps.
    \item Enable application access security by enforcing rules-based Multi-Factor Authentication for both on-premises and cloud applications.
    \item Provision secure remote access to on-premises web applications through Azure AD Application Proxy.
\end{enumerate}

\begin{figure}[htp]
    \centering
    \includegraphics[width=15cm]{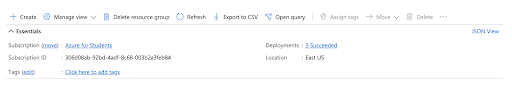}
    \caption{Azure Identity Management from Azure Portal}
    \label{fig:galaxy}
\end{figure}

% ############################

\textbf{TR31 (BR14) -  The system should allow usage of scripts for the operations procedures and automate their execution by triggering them in response to events.}

\begin{figure}[htp]
    \centering
    \includegraphics[width=10cm]{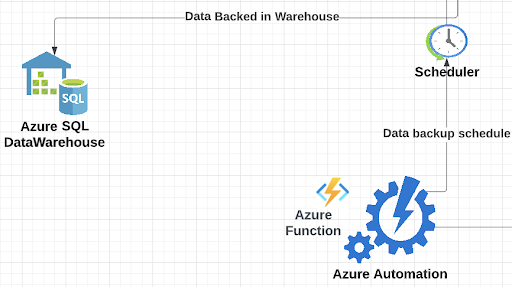}
    \caption{Azure Automation using Azure Functions}
    \label{fig:galaxy}
\end{figure}

\textbf{Automation} services can define input, action, activity to be performed, conditions, error handling, and output generation [14]. Certain services allow users to schedule a variety of tasks or move their focus from manually conducting operational duties to creating automation for these operations, including:
\begin{enumerate}
    \item Reduce time to perform an action.
    \item Reduce risk in performing the action.
    \item Increased human capacity for further innovation.
    \item Standardize operations.
\end{enumerate}

\textbf{Azure Functions} has script that enables a recurring backup of data from Azure Data Lake to Azure SQL Data Warehouse is contained in the Azure Function for this system. The mechanism for batch processing the data is also incorporated in the script, and this occurs once every 12 hours. The scheduler-connected automation service is triggered by the function. Additionally, the script can be changed to carry out additional tasks like starting or shutting down idle virtual machines. The overall cost and resource use could be optimized in this way. Thus, using scripts enables standardization of operational processes and execution of triggers in response to events, satisfying requirements.

% ###########

\textbf{TR61 (BR24) - The infrastructure should be secured through a VPC, having isolation of physical hosts and controlling network traffic.}

\begin{figure}[htp]
    \centering
    \includegraphics[width=10cm]{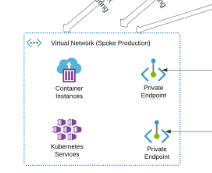}
    \caption{Azure Automation}
    \label{fig:galaxy}
\end{figure}

By default, a \textbf{VNet}'s resources can communicate outbound to the internet. It is possible to extend the private address space and identity of the virtual network to Azure service resources through a \textbf{virtual network service endpoint} [16]. Systems can link virtual networks to one another through virtual network peering, allowing resources in either virtual network to communicate with one another across multiple regions.

According to our architecture, the system is separated into a number of components, each of which is located within a virtual network space. Virtual networks are used to separate containers and kubernetes, connect with trusted and untrusted users, and provide services like databases, virtual machines, automation, and monitoring. This validates the criteria by ensuring physical host isolation and network traffic control.

% #######################

\textbf{TR48 (BR20) - The system should enable firewall settings to restrict access to the infrastructure for the untrust users.
}

\begin{figure}[htp]
    \centering
    \includegraphics[width=10cm]{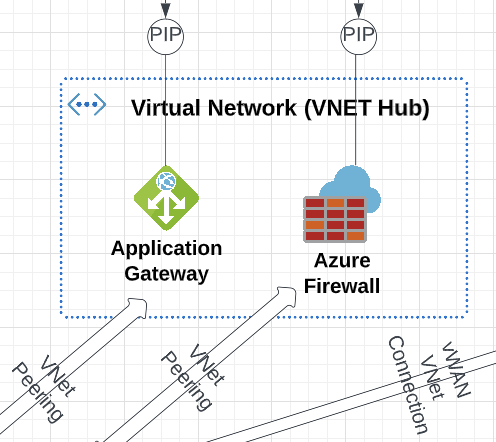}
    \caption{Azure Firewall}
    \label{fig:galaxy}
\end{figure}

Azure Firewall has some important features including:
\begin{enumerate}
    \item Availability Zones
    \item Application FQDN filtering rules
    \item Network traffic filtering rules
    \item Threat intelligence
    \item DNS proxy
    \item Multiple public IP addresses
    \item Certifications
\end{enumerate}

\textbf{Azure Firewall} is fully stateful, so it can distinguish legitimate packets for different types of connections [18]. The firewall can be configured to use threat intelligence-based filtering to warn about and block traffic from/to known malicious IP addresses and domains. Azure Firewall may process and forward DNS requests from a Virtual Network(s) to the chosen DNS server when DNS proxy is configured.

Based on the firewall DNS settings, the provided FQDNs in the rule collections are converted to IP addresses. Since this function relies on DNS resolution, it is strongly advised to turn on the DNS proxy to guarantee that name resolution is consistent with the firewall and protected virtual machines. The design successfully meets our requirements, making the firewall a crucial system component.

% #########################

\textbf{TR45 (BR19) - Cloud Networking WAN should be separated as TWO Virtual Networking Environments, one for trusted users and another for untrust users.
}
\begin{figure}[htp]
    \centering
    \includegraphics[width=10cm]{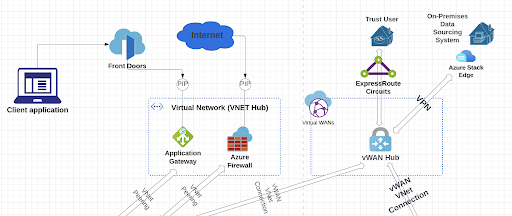}
    \caption{Azure VNet / VWaN for Trust and Untrust Users}
    \label{fig:galaxy}
\end{figure}

A secure internet tunnel is used to transmit communications between the client computer and the virtual network. Any on-premises resource that the system permits can access a virtual network using this connection type. Through an ExpressRoute partner, an Azure ExpressRoute is established between the network and Azure. The traffic does not cross the internet on this private connection [17].

A network virtual appliance is a VM that handles a particular network task, like WAN optimization or a firewall. When used together, vWANs, Express Route, and Front Doors[19] provide excellent safety and security to separate requests from users who are trusted and those who are not in order to fulfill the requirements.

% ##################

\textbf{TR28 (BR13) - The system should ensure that the backed up data is encrypted when in flight and when at rest.}

Authorization manages access rights for Data Lake Storage Gen1 when Azure Active Directory authenticates a user so that the user can use Data Lake Storage Gen1.

Additionally, Data Lake Storage Gen1 offers encryption for the information kept in the account. Data stored in Data Lake Storage Gen1 is encrypted before being stored on persistent media if encryption is chosen for the data. In this scenario, Data Lake Storage Gen1 automatically encrypts data before persisting and decrypts data before retrieval, all of which is invisible to the client requesting access to the data. 

\begin{figure}[htp]
    \centering
    \includegraphics[width=10cm]{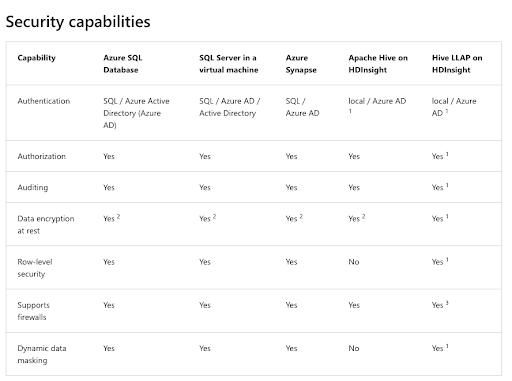}
    \caption{Data Security Capabilities}
    \label{fig:galaxy}
\end{figure}

%  ###################

\textbf{TR19 (BR10): The system should have a 99.999\% throughput with a maximum acceptable latency under 3 seconds.}

\begin{figure}[htp]
    \centering
    \includegraphics[width=10cm]{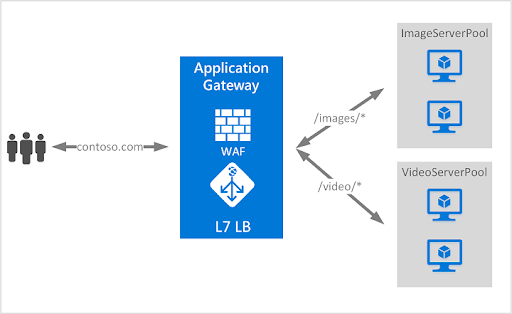}
    \caption{API Gateway and Load Balancer}
    \label{fig:galaxy}
\end{figure}

The most effective load balancer for Azure is Azure Load Balancer, which also maintains extremely low latency. The inbound traffic flows can be divided between a load balancer's frontend and backend pools by establishing a load balancer rule. A distribution method called \textbf{session persistence}, also referred to as session affinity, routes to backend instances using a two- or three-tuple hash of the source IP, destination IP, and protocol type[22]. The same client's connections will all route to the same backend instance in the backend pool when session persistence is used.
High performance and throughput from the system, as well as the use of API Gateways (that enable Reverse Proxy and SSL Certification Resolution), ensuring that the system is available 99.999\% of the time, meeting the requirements.

% ####################

\textbf{TR58 (BR22) - The system should allow monitoring the amount of requests being served and data being flown into the system.}

\begin{figure}[htp]
    \centering
    \includegraphics[width=10cm]{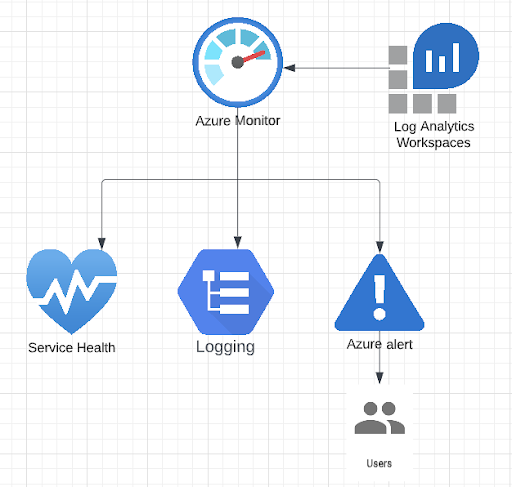}
    \caption{Azure Monitor Setup}
    \label{fig:galaxy}
\end{figure}

Based on the namespaces of the resource providers, \textbf{Azure Monitor}[23] data is gathered and saved. All unique IDs include the namespace of the resource provider. A resource ID for a key vault might like /subscriptions/subscription\_id/resourceGroups/.Each of the following layers is where Azure Monitor gathers data from:
\begin{enumerate}
    \item \textbf{Application} - Data about the performance and functionality of the code written.
    \item \textbf{Container} - Data about containers and applications running inside containers, such as Azure Kubernetes.
    \item \textbf{Guest operating system} - Data about the operating system on which the application is running. 
    \item \textbf{Azure resource} - Data about the operation of an Azure resource. 
    \item \textbf{Azure subscription} - Data about the operation and management of an Azure subscription, and data about the health and operation of Azure itself.
    \item \textbf{Azure tenant} - Data about the operation of tenant-level Azure services, such as Azure Active Directory.

\end{enumerate}

By proactively informing service users through the data from Azure Monitor services, alerts help to identify and address issues before users notice them. The alert rule intercepts the signal and determines whether it satisfies the condition's requirements. If the criteria are satisfied, an alarm is raised, starting the related action group and updating the alert's status.

\begin{figure}[htp]
    \centering
    \includegraphics[width=10cm]{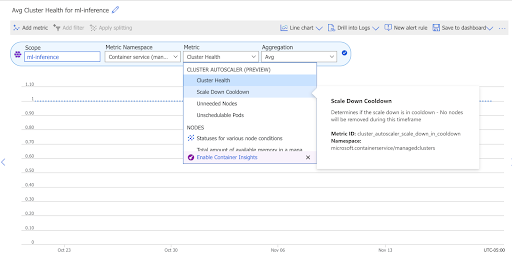}
    \caption{Azure Monitor Dashboard from Azure Portal}
    \label{fig:galaxy}
\end{figure}

The Azure Kubernetes Service (AKS), which houses the ML-Inference engine for our application, is monitored for its health and performance through Azure Monitor services. Container insights offers spreadsheets and interactive views that evaluate gathered data for various monitoring scenarios.
Along with Azure Monitoring, Azure Alerts, and Logging tools, the planning and configuration of Azure Log Analytics helps fulfill the business and technical needs of monitoring the requests and data flow for each of the system's resources.

% ########################

\textbf{TR12 (BR6): The system should enable the usage of appropriate services, resources and configurations for the workloads to optimize the overall cost spent on the services. }

When Azure compute resources are regularly used, Azure savings plans generate savings. By allowing one to commit to a predetermined hourly spend on computing services for one or three years, an Azure savings plan facilitates cost-saving measures[24]. By as much as 65\% compared to pay-as-you-go charges, a savings plan can drastically lower the cost of resources. Instead of commitment amount, discount rates per meter vary by commitment length (1-year or 3-year). Up until the system achieves the commitment amount, compute usage is discounted hourly under a savings plan; any more usage after that is charged at pay-as-you-go rates. 

It would be advisable to purchase a savings plan in order to lower total costs because training and inference engines use a lot of processing time and compute power. Purchases made with a savings plan cannot be reversed or reimbursed.

Customers can deploy and operate discounted Windows Server virtual machines in Azure using local Windows Server licenses protected by Software Assurance under the A-HUB (Azure Hybrid Use Benefit) plan[25]. Then, customers simply have to pay the fundamental virtual machine fees, which are the same as those for Linux virtual machines.

\begin{figure}[htp]
    \centering
    \includegraphics[width=10cm]{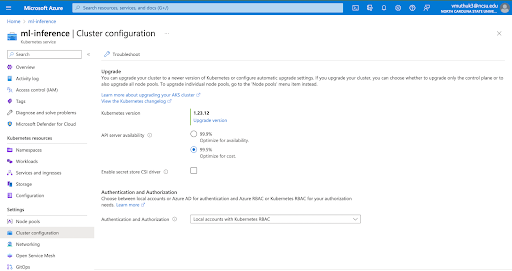}
    \caption{Cost Configuration of service from Azure Portal}
    \label{fig:galaxy}
\end{figure}

% ############################

\textbf{TR34 (BR15) - The system should scale out by one instance if average CPU usage is above 70\%, and scale in by one instance if CPU usage falls below 50\%.
}

The system can specify a specific number of resources to employ to maintain a fixed cost, such as the number of nodes, by manually scaling resources[26]. The replica or node count is set for manual scaling. Based on that replica or node count, the Kubernetes API then schedules building additional pods or draining nodes. The number of underlying Kubernetes nodes may also need to change as the number of application instances does.

The horizontal pod autoscaler (HPA) is a tool that Kubernetes employs to monitor resource consumption and scale the number of replicas automatically. The Metrics API collects data from the Kubelet every 60 seconds, whereas the Horizontal Pod Autoscaler checks the Metrics API every 15 seconds for any required adjustments in replica count. The HPA is actually updated every 60 seconds. The quantity of replicas is adjusted upwards or downwards as necessary.

\begin{figure}[htp]
    \centering
    \includegraphics[width=10cm]{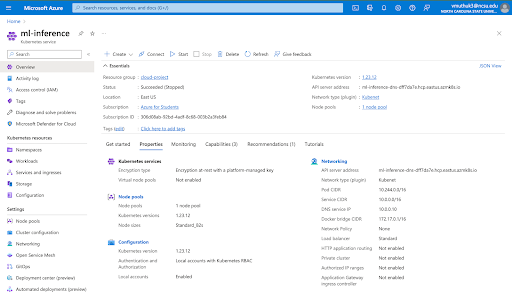}
    \caption{Azure Kubernetes Configuration from Azure Portal}
    \label{fig:galaxy}
\end{figure}

The minimum and maximum number of replicas that can operate can be set while setting the horizontal pod autoscaler for a specific deployment. Additionally, the provider can specify the statistic to track and use as the basis for any scaling decisions, such as CPU utilization. This behavior enables the Metrics API to represent the distributed workload and the updated replica count to go into effect.

% ######################
\subsection{Design principles and best practices used} 
The system architecture is designed by following some of the best design practices and principles.
\begin{enumerate}
    \item \textbf{Design for automation} - Regarding the particular application that is developed on the cloud system, MLOPS would call for the use of numerous resources in order to periodically feed in data, execute preprocessing, create inference engines, store and backup data, etc. Automation is crucial in keeping a system capable of handling such a large number of data, processing, and requests together.
    \begin{enumerate}
        \item \textbf{Continuous Integration and Continuous Delivery}: A periodic and iterative release of the code with structured changes is highly significant since the system uses code and scripts to do data preprocessing, preparation, and model deployment. This release is controlled by CI/CD environments like Jenkins, which are integrated with our system.
        
        \item \textbf{Scale up and Scale down}: A resilient system needs a number of key design principles, including autoscaling and flexibility. The Azure Kubernetes Service (AKS), which scales up and down the cluster based on the CPU load on the inference engines that compute huge models, is used in this architecture. This promotes efficient resource allocation and use, which reduces costs significantly.
        
        \item \textbf{Logging and Monitoring}: The system incorporates logging and monitoring of each service's actions to analyze service level performance. Every request and process is additionally recorded to aid with business-level problems. The proper stakeholders are informed of the problems at the appropriate time through the integration of alerting and tracking technologies. They can provide insightful information about user behavior and system usage (such as how many individuals are using the system, which components they are utilizing, and how late they are on average). Additionally, they can be used collectively to calculate the overall system health.
    \end{enumerate}

    \item \textbf{Practice Defense in Depth}: Every single piece of information in the cloud must be protected and securely encrypted because users choose to use cloud systems and infrastructure to store their data, conduct computations, and respond to requests. Because cloud-native architectures started as internet-facing services, they have always had to defend against outside threats. Encrypting the data while storing and maintaining backups is crucial because the system deals with massive volumes of data. It's also a good idea to make the request endpoints private. Every tenant's login information and other crucial business information shouldn't be compromised. Hacker threats and assaults ought to be prevented.

    \begin{enumerate}
        \item \textbf{Authentication of each component} : The Active Directory, which authenticates users accessing the service, must be updated to include each and every component that is registered as a part of the service.
        
        \item \textbf{User level authorization} : The trust and untrust users are identified separately and they are filtered through multiple levels of security through firewall, front door, api gateway, express routes etc before the request reaches the specific services. Multi factor authentication should be enabled for every user.
        
        \item \textbf{Rate Limiting and Prevent Script Injection} : Every service has a cap on how many requests it will accept from a user in a particular period of time. If the system receives an abnormally high volume of queries, they could be suspected of being malicious. The code that is often attached in message headers and body as encrypted packages cannot be allowed to contain external code or script.

    \end{enumerate}

\end{enumerate}

\subsection{Tradeoffs revisited}
\label{Tradeoffs revisited}

The system was architected by handpicking a few technical requirements that solve major of the business requirements. Each technical requirement and the corresponding business requirements were translated into actionable systems using the services provided by the cloud service provider. This makes every component of the system robust to satisfy the specific requirements. However, when they all collaborate to form a large functional system in a real world scenario, there needs to be multiple tradeoffs done to analyze the actual successful implementation of the system. Some of the important tradeoffs from our system architecture are discussed in detail below:

\begin{enumerate}
    \item \textbf{Security vs Performance Efficiency Tradeoff}: It is important to have the system secured from the internal services as well as from the external users. To enable these, there has to be multiple levels of encryption and security checkpoints that have to be detailed out in the system. Having so many layers of security naturally affects the response time of the system and thereby not being able to meet the 99.999\% performance efficiency in terms of throughput and availability. In terms of the requirements, the conflicting technical requirements for this scenario is:
    \begin{enumerate}
        \item \textbf{TR19 (BR10)} - The system should have a 99.999\% throughput with a maximum acceptable latency under 3 seconds.
        \item \textbf{TR48 (BR20)} - The system should enable firewall settings to restrict access to the infrastructure for the untrust users.
    \end{enumerate}

    Firewall restricts the access from the external users of the system who are also known as untrusted users. The firewall uses a threat intelligence-based filtering can to alert and deny traffic from/to known malicious IP addresses and domains. This protects the infrastructure from untrusted users outside of the scope of the system and prevents unwanted attacks. With DNS proxy enabled, Azure Firewall can process and forward DNS queries from a Virtual Network(s) to the desired DNS server by resolving the DNS names. All of these processes naturally take a few milliseconds of time to make the request reach the system. In addition the system might have its own processing time and thereby causing the throughput to reduce drastically.

    \begin{enumerate}
        \item \textbf{TR61 (BR24)} - The infrastructure should be secured through a VPC, having isolation of physical hosts and controlling network traffic. 
        \item \textbf{TR19 (BR10)} - The system should have a 99.999\% throughput with a maximum acceptable latency under 3 seconds.
    \end{enumerate}

    Looking at the requirements to secure every host / service through a VPC, there would be private endpoints required to reach every service which resides inside a VNet. This would in turn cause some amount of latency.

    However the latency is not that alarming to affect the performance of the whole system. But a security threat is more vulnerable for the system, the users and the business applications running on the system, Hence it is highly necessary to secure the system to the best possible way making the tradeoff with a really high throughput ratio.

    \item \textbf{Reliability vs Performance Efficiency}: This is another critical tradeoff to be looked at while designing the system. Just like the system has to be secure, the system should also be reliable to faults, disasters and have a very sophisticated way of recovering the infrastructure, data and other compute resources. To integrate a pipeline / process workflow for having a reliable system along with their computational performance, a lot of automation and backup services have to be linked to the main compute services. These services would have to address the requests for backup along with the actual computational requests received from the external users. Hence the overall latency for delivering the response for these requests increases which causes a less than expected performance efficiency. Looking at the specific pair of requirements where the tradeoff has to be made:
    \begin{enumerate}
        \item \textbf{TR19 (BR10)} - The system should have a 99.999\% throughput with a maximum acceptable latency under 3 seconds.
        \item \textbf{TR28 (BR13)} - The system should ensure that the backed up data is encrypted when in flight and when at rest.
    \end{enumerate}
    
    These requirements specifically focus on having the data backed up and the data should also be encrypted while they are stored and while they are in motion. The actual data storage would happen with the big data service called the Azure Data Lake. Once the data is in place for processing, an automation system sends out periodic requests to move the data into a backup system (Azure SQL Data Warehouse). To ensure the data is protected in transit, the data is encrypted before it is sent to the backup service. Also the end point accessing the services is made private. Once it reaches the backup server, the api is resolved and the encrypted data from the request body is recovered. Then the data is again encrypted as per the standards of SQL to store it in the backup database.

    This level of periodic backup and ensuring the protection of data makes the system extremely robust and reliable to threats. However the repeated encryption and decryption affects the latency and throughput of the system. Prioritizing the reliability of the system, the tradeoff has to be taken for compromising on the throughput and efficiency. Hence the emphasis in design was given for having a more reliable and robust system.

    \item \textbf{Cost vs Operational Efficiency Tradeoff}: Cost tradeoff has always been very important while designing and implementing any system. Since the primary emphasis goes on reducing the cost, the architecture would end up choosing low cost services, resources and configuration thereby sacrificing the operational efficiency of the system.
    \begin{enumerate}
        \item \textbf{TR12 (BR6)} - The system should enable the usage of appropriate services, resources and configurations for the workloads to optimize the overall cost spent on the services. 
        \item \textbf{TR31 (BR14)} -  The system should allow usage of scripts for the operations procedures and automate their execution by triggering them in response to events.
    \end{enumerate}
    
    Comparing the requirements for this system in terms of the operational efficiency and cost, it is important to have a system being automated for tasks like risk monitoring, alerting, periodic backups, scheduled system start / stop and system maintenance etc. Enforcing all of this would make the system perform all activities smoothly. However, this will make the usage of additional resources which in turn increases the cost of the overall system. Since, the operational efficiency is to be prioritized for the efficient functioning of the system, we choose to have the automation services in place taking the tradeoff against the cost.

    \item \textbf{Cost vs Performance Efficiency Tradeoff}: The performance of the system is of the highest priority while designing the architecture. To support the high availability, throughput and low latency of the system, it is important to have a good monitoring system to keep track of the requests, load and other performance metrics of each of the services.
    \begin{enumerate}
        \item \textbf{TR12 (BR6)} -  The system should enable the usage of appropriate services, resources and configurations for the workloads to optimize the overall cost spent on the services.
        \item \textbf{TR58 (BR22)} - The system should allow monitoring the amount of requests being served and data being flown into the system.
    \end{enumerate}
     
     Comparing the technical requirements, employing monitoring resources to track the performance for each system, as well as logging the data and requests for business level troubleshooting requires extra configuration at a service level. Also this will require the usage of more resources to store the logs generated, perform analytics of the log and take backup of the whole monitoring infrastructure. Hence, this boils down to a similar tradeoff of shelling extra money for improving the efficiency of the system. While the user’s primary focus would be to have a highly available, there will be a definite tradeoff between the cost and performance efficiency. Hence performance efficiency is given more importance over the cost by employing additional resources and configurations for optimal workload generation.

\end{enumerate}

\newpage

\section{Kubernetes experimentation}
\label{sec:Kubernetes experimentation}
%% YP: What do we mean by Kubernetes experimentation? Are we going to enforce that students use Kubernetes?
%%YV: since k8s is the only orchestration tool we'll discuss in the lectures, yes. I am open to alternatives, however.

\subsection{Experiment Design} 
The technical requirement that we choose to work with in this section is mentioned below:

\textbf{TR34 (BR15) - The system should scale out by one instance if average CPU usage is above 70\%, and scale in by one instance if CPU usage falls below 50\%.}

To achieve this requirement, we plan to make use of \textbf{Azure Kubernetes Service} and configure the application, clusters and nodes accordingly.

The \textbf{Training cluster} and the \textbf{ML inference cluster} are the two Kubernetes-related components of our planned solution. We will focus more on ML inference cluster design in this section. Probability is used to compute the ML model's output, making it challenging to assess the efficacy of each iteration. Murphy's law states that "Anything that can go wrong will go wrong." As a result, our attention will be on Istio-based Canary Deployments.

\textbf{Canary deployments} are a method of staged release [12]. To test the upgrades and get user input, we release them to a select sample of users first. The remaining users will receive the update after the modifications are approved. Canary deployments might be accomplished extremely well with Service Mesh (Istio).

The most well-known open source service mesh software is called \textbf{Istio}. Service-to-service communication is handled by a separate infrastructure layer called a service mesh. It is in charge of ensuring that requests are delivered reliably through the intricate web of services that makes up a contemporary, cloud-native application. In actual use, the service mesh is usually implemented as a collection of thin network proxies that are installed concurrently with application code without the requirement for the application to be aware of it.

\begin{figure}[htp]
    \centering
    \includegraphics[width=10cm]{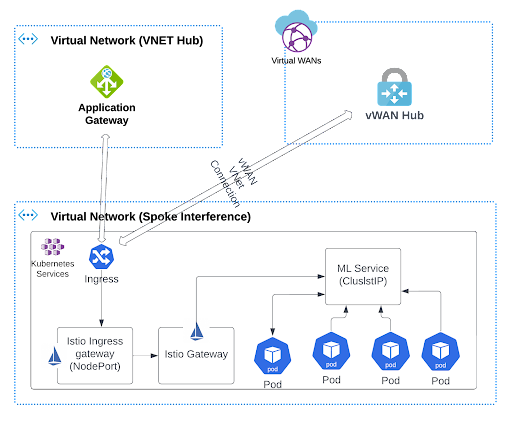}
    \caption{Kubernetes System Architecture}
    \label{fig:galaxy}
\end{figure}

\textbf{\underline{Working of the Architecture}}

Both trusted and untrusted users will connect to Ingress, which exposes HTTP and HTTPS routes to services inside the cluster from outside the cluster [11]. Ingress will establish a connection with the Istio Ingress gateway, a NodePort for exposing services on a predetermined port. At the mesh's edge, Istio Gateway is a load balancer that accepts and routes HTTP/TCP connections [11]. Istio Gateway is connected to our ML Service, a ClusterIP that offers network communication within the ML cluster, as a final step. Normally, it can only be used for internal networking among your workloads and cannot be accessed from the outside. Every iteration of the ML model will be a pod.

\textbf{\underline {Canary deployments}}

Prior to discussing the canary deployment, the definitions of VirtualService, Destination Rule, and deployment/service should be given.

Istio's traffic routing capabilities is based on two fundamental building blocks: \textbf{Virtual services} and \textbf{Destination rules}. Request routing to a service within an Istio service mesh could be customized by a \textbf{virtual service}. A set of routing rules makes up each virtual service, allowing Istio to match each virtual service request to a particular physical destination inside the mesh [11].

Istio's traffic routing capabilities also includes \textbf{Destination rules}. The virtual services determine how to route the traffic to a specific destination, and then it may control what happens to that traffic using destination rules. We categorize the machine learning services in our instance according to version.

\begin{figure}[htp]
    \centering
    \includegraphics[width=10cm]{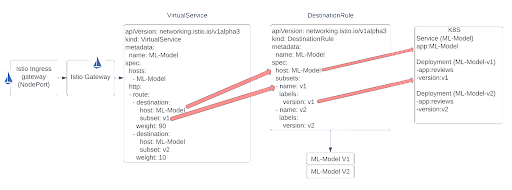}
    \caption{Canary Deployments with Istio}
    \label{fig:galaxy}
\end{figure}

\begin{enumerate}
    \item The \textbf{“host:ML-Model”} in the VirtualService configuration corresponds to the “host:ML-Model” in the destination rule configuration.
    
    \item \textbf{“subset:v1”} in the VirtualService configuration corresponds to “subsets: -name:v1” in the destination rule configuration.
    
    \item The \textbf{"host:ML-Model"} in the destination rule configuration corresponds to the service "ML-Model" in k8s. In k8s, and corresponds to the service in the istio registry that the rule acts on.
    
    \item \textbf{"labels: version:v1"} in the destination rule configuration corresponds to "- version:v1" in the deployment in Kubernetes. The corresponding subset1 forwards the traffic to the service corresponding to the deployment with the label version:v1.
    
    \item In the VirtualService setup, we only move 10\% of traffic to ML-Model v2 and maintain 90\% of traffic on ML-Model v1.
\end{enumerate}

% \textbf{\underline{The Kubernetes Cluster configuration from the Azure Portal is done as shown - insert pic from azure portal}}

\subsection{Workload generation with Locust}
    
We developed a Python script file for locust to produce the workload. We used the task tag within the code to identify the multi-threaded running tasks for API queries and payload-sending logic within the script file.

The task scheduler logic is then invoked, and the endpoints. We have chosen to employ ramp-style user spawning logic to gradually increase users in 20 seconds with 20 users, implying that the spawn rate would be one user per second, in order to set the stage for the stress testing.

At a spawn rate of 100, 6000 users will be generated between 20 and 80 times every second. The number of users will decrease by 100 users per decade from 80 to 140, reaching 20.

The final stage ends the show completely after 1140 to 160 seconds and reduces the number of clients to 10 at a drop rate of 0.5 clients per second.

\begin{figure}[htp]
    \includegraphics[width=10cm]{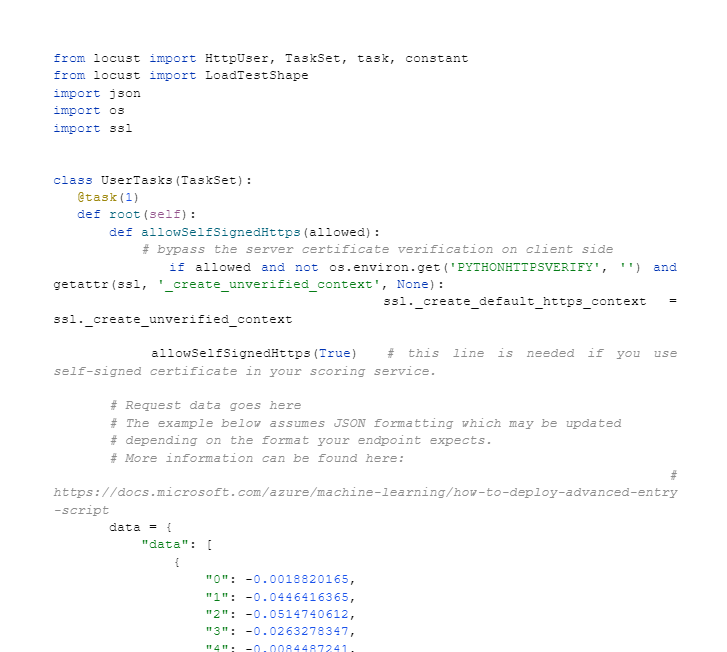}
    \caption{Code Snippet to Initiate Parallel Request Loads}
    \label{fig:galaxy}
\end{figure}

\begin{figure}[htp]

    \includegraphics[width=10cm]{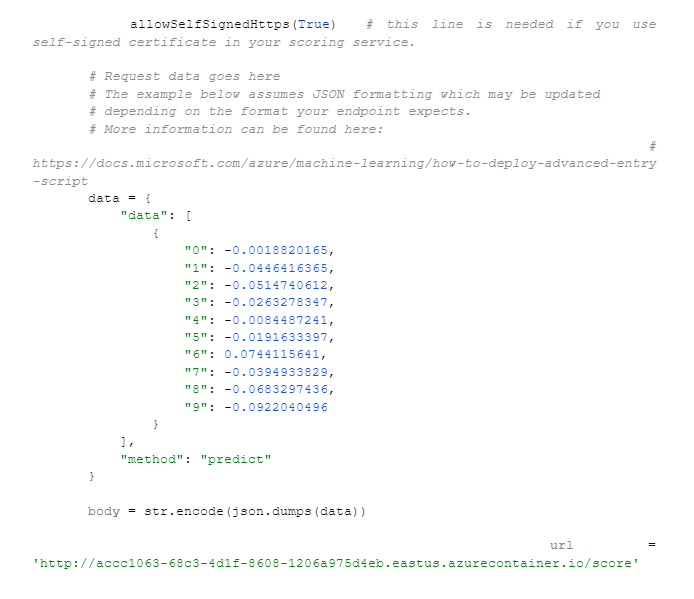}
    \caption{Data Payload for the Inference Engine}
    \label{fig:galaxy}
\end{figure}

\begin{figure}[htp]

    \includegraphics[width=10cm]{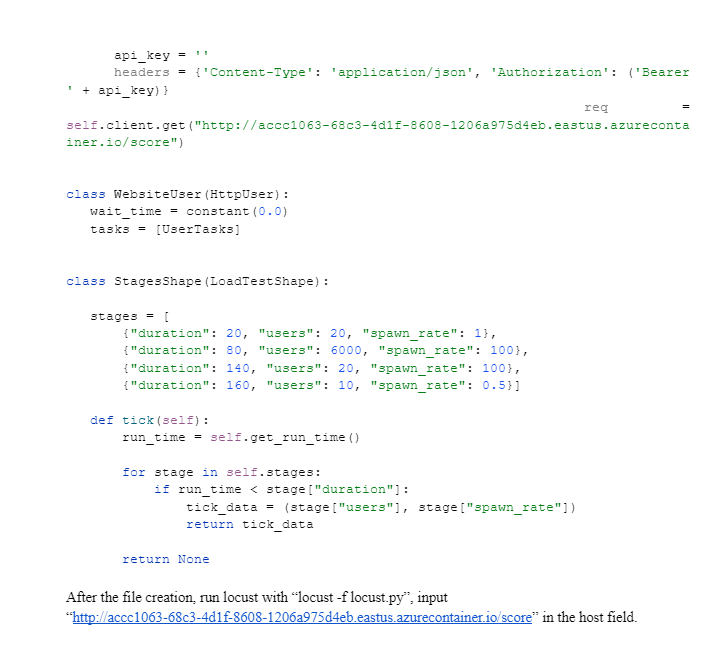}
    \caption{Defining Ramp and Number of Users}
    \label{fig:galaxy}
\end{figure}

% #################### 
\subsection{Analysis of the results} 

\begin{figure}[htp]
    \centering
    \includegraphics[width=15cm]{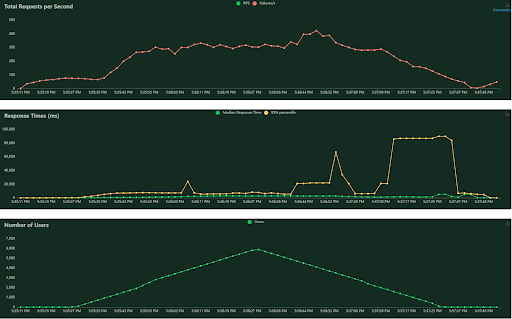}
    \caption{Load Generation using Locust}
    \label{fig:galaxy}
\end{figure}

We employed a progressive gradient ramp for the user spawn pattern in the load testing operations. The graph also seems to have a particular local network bottleneck that is the cause of the odd trend in the total requests per second.

We can see from the Locust load generation analysis result graph above that as user counts rise, the system response time latency increases, however it does so more slowly.

As can be seen, the response time graph has a significant cliff. Given the fact that there appear to be several smaller peaks both before and after the maximum user count, we can deduce that the architectural design of the cloud computing system required more time to develop than anticipated.

Overall, the open endpoint's availability and rising resource use as the number of concurrent user requests per second increased served to validate the architecture.

The delay also demonstrates the need for further design modifications to handle edge-case scenarios.

The flooding protection service intervenes to stop the abuse of cloud computing resources, safeguarding both the service of the cloud consumer and the resources of the cloud provider, as seen on the graph, following a brief period of time in which repeated requests are made to the same endpoint.
\newpage

\section{Comparisons}
\label{sec:Comparisons}

\begin{figure}[htp]
    \centering
    \includegraphics[width=15cm]{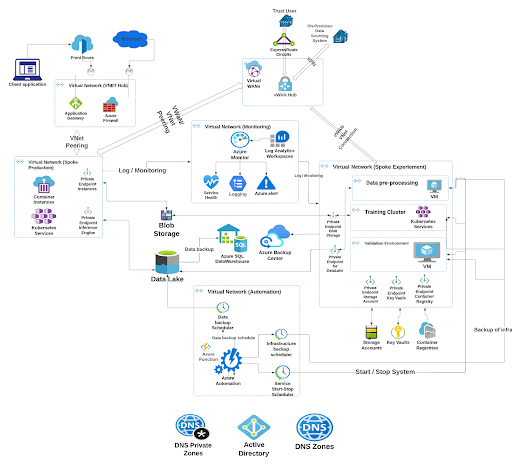}
    \caption{Extended design diagram with more services covering other data paths.}
    \label{fig:galaxy}
\end{figure}

This section will compare two approaches to developing an architecture with more services introduced in accordance with modern trends and trace the entire data path of each service's operation through user requests. The \textbf{Monitoring} VNet, for example, monitors performance, combines logs, and establishes warning mechanisms for the core SPOKE production and experiment services. This expanded solution makes use of additional modules in the system architecture. In order to guarantee uptime and lower system costs, services are also made available for \textbf{automating} certain operations like data backup, backup of infrastructure, and periodic start and stop of services. There are other deployed \textbf{backup services} like Azure SQL Data warehousing and Azure Backup Centers.

The various architectural elements are further compared and explained in this part in terms of the services offered and the data flow channel.

\begin{enumerate}
    \item \textbf{Request flow path between TRUST vs UNTRUST systems}.

    Data is being ingested from the on-premise provider system. Since they are a member of the system, these users are also referred to as the \textbf{Trust Users}. The \textbf{Azure IOT Edge} is integrated into the devices, and \textbf{ExpressRoute} is used to broadcast the data without using the internet ensuring data protection. \textbf{Untrusted Users} refers to the external client application that uses the internet to access the system. The requests coming from these users must be examined, and the procedures for \textbf{Firewall} and \textbf{Front Door} authentication and authorisation must be used. As a result, excellent security is ensured and unwelcome attacks from outside dangers are avoided.

    \begin{enumerate}
        \item \textbf{VWan vs Hub-Spoke}

        A virtual network called the \textbf{Hub} serves as a focal point for connectivity to the on-premises network in Azure. Workload isolation can be accomplished using the \textbf{Spokes}, which are virtual networks that communicate with the hub. The hub and the on-premises data centers are connected by an ExpressRoute or VPN gateway, which allows traffic to pass between them. The primary difference of this strategy is the usage of managed \textbf{Azure Virtual WAN (VWAN)} in place of hubs. The typical hub-spoke network topology's advantages are present in this architecture, along with several new advantages:
        \begin{enumerate}
            \item By substituting existing hubs with a fully managed VWAN solution, operational overhead is reduced.
            
            \item Using a managed service and doing without network virtual appliances results in cost savings.

            \item Reduced security concerns associated with misconfiguration by establishing centrally controlled secured Hubs with Azure Firewall and VWAN.
            
        \end{enumerate}

    \end{enumerate}

    \item \textbf{Compare Open Source vs Cloud based API Gateway and Load Balancing}

    To route requests from outside clients while getting to the inference environments, the system offers an Azure API Gateway and Azure Load Balancer. Due to the huge volume and fact that these requests are transmitted via the internet, it is crucial to secure them using SSL certificates and prevent direct server exposure. An API Gateway would use reverse proxies and resolve the SSL certificates in this situation. The requests are then served by a load balancer set up in one of the methods (round robin, FCFS, etc.). These cloud-based load balancers offer a wide range of services, functions, and benefits in terms of performance and scalability when compared to contemporary open source load balancers like Facebook Katran.

    \begin{enumerate}
        \item \textbf{Facabook Katran vs Azure Load Balancer}

        \textbf{Cloud load balancers} are reasonably priced, and usage-based fees apply. AWS, GCP, and Azure's closed source load balancers are quick and dependable, but they offer little room for customization and risk lock-in. On the other hand, \textbf{open source load balancers} offer great performance and a variety of deployment choices, including support for both cloud and on-premise deployments. Popular open source software initiatives additionally benefit from active user communities.
        
        By utilizing two recent advancements in kernel engineering, the \textbf{eXpress Data Path (XDP} and the \textbf{eBPF virtual machine}, \textbf{Katran} [10] provides a software-based approach to load balancing with a completely redesigned forwarding plane. They made the decision to develop an open source program that is hardware-optimized in order to manage the amount of requests and the pace of operations at Facebook's backend services. The first generation of L4LB was based on the principles of :
        \begin{enumerate}
            \item VIP announcement
            \item Backend server selection
            \item Forwarding plane
            \item Control plane
        \end{enumerate}

        To prevent having to compute the hash twice for subsequent packets, each L4LB also maintains the backend selection for each 5-tuple as a lookup table. But having the L4LB and a backend on the same host raised the risk of a device failing, which was a drawback. Then, a second-generation L4LB with a fully redesigned forwarding plane was developed [10].

        \begin{enumerate}
            \item Working in tandem with the Linux networking stack, the XDP offers a quick, configurable network data channel without using a full-fledged kernel bypass mechanism.
            
            \item In order to interface with the Linux kernel and increase its functionality by executing user-space supplied programs at specific locations in the kernel, the eBPF virtual machine offers a flexible, effective, and more dependable method [10].
        \end{enumerate}

        \textbf{\underline {Comparing open source and proprietary cloud based load balancers}}

        \begin{enumerate}
            \item \textbf{Responsiveness} - Tools that are open source are supported, maintained, and worked on by a sizable development community. They respond quickly to problems and settle them once they arise. But the exclusive cloud-based ones take their time to provide a solution.

            \item \textbf{Flexibility} - The inherent adaptability of open source load balancers allows for quick and simple compliance with various regulations. Proprietary platforms, on the other hand, might be difficult to adapt, making vendors less able to quickly alter their goods to suit shifting client demands.

            \item \textbf{Support} - Customers still select cloud-based load balancers for the most part for their broad support and widespread usage. The cloud vendors have established guaranteed service level agreements (SLAs). No SLA has been created by the open source community. While some, like HAProxy and Katran, are developed and might survive in this context, others may lose the interest of developers, and might disappear altogether.
        \end{enumerate}
    \end{enumerate}
\end{enumerate}

\newpage

\section{Conclusion}
\label{sec:conclusion}
 
\subsection{The lessons learned}

% {\color{red} Describe, in detail, your experience with this project.} What was difficult? interesting? boring? unexpected? etc.

One of the interesting things we discovered while working on the project is that the technical requirements were more complicated than we had anticipated because, as we composed the technical requirements, we also needed to take the business requirements and the principles of Well Architected Framework into consideration. We need delve deeply into the well-designed architecture framework's specifics and adhere to numerous principles across various aspects of consideration, such as security and performance effectiveness, to ensure that the technical requirements are contained within the framework.

We also thought the brainstorming process used to select the topic to be interesting. The team came up with numerous ideas for moving web applications to the cloud and other local applications to the cloud during the project's first brainstorming session. The team has chosen to employ a bottom-up strategy rather than a top-down strategy to take into account the scope of improvement in a future application.

We came to the realization that the trade off is a key consideration when deciding which characteristics of the technical requirements should be translated from the business requirements. In comparison to the traditional software solution that operates locally, cloud computing solutions have higher technical requirements in terms of factors like cost, security, and application performance.

As was already mentioned, one unexpected aspect was how taking into account the Well Architectural Framework allowed for a different perspective to be looked into the design choices from various visions and hats across the solution when attempting to link technical requirements to business requirements.

The team was also startled by the complexity of the system's configuration because it uses Azure as a cloud provider and it takes more time to deploy and implement the cloud application's existing solution utilizing Azure native services that are housed in the cloud.

\subsection{Possible continuation of the project}

% If you plan to take the advanced course during the spring 2023 semester, {\color{red} provide here some ideas on how to continue the project.} Ditto if you are interested in further research/publications. 

One of the few future implementations taken into account is further Kubernetes development, including grey release and design consideration. There are also more options for study on things like additional design details with regard to both business and technical requirements. In order to thoroughly examine the trade-off between various cloud providers, it is also possible to provide a more thorough comparison of cloud providers.

\newpage

\section{References}
\label{sec:references}

% {\color{red} Include here your references.}

% Reference \cite{reference1} is a dummy reference to show you how you can supply a reference using the \\
% \verb"\bibitem{myreference}" \\
% latex command; once you have entered the command, you can cite the reference in your text using a \\
% \verb"\cite{myreference}" \\
% command. See \cite{reference2}.

%\input{}\input{}\input{}\input{}\input{}

\end{document}